\documentclass[preprint,aps,a4paper,superscriptaddress]{revtex4}
\usepackage{graphicx,amsmath,amssymb,psfrag,epsfig,multirow,slashbox,bibentry,verbatim,color,epstopdf,comment}
\begin{document}
\title{Modeling the long term dynamics of pre-vaccination pertussis}
\author{Ganna Rozhnova\footnote{corresponding author\\a\_rozhnova@cii.fc.ul.pt}}
\affiliation{Centro de F{\'\i}sica da Mat\'eria Condensada and 
Departamento de F{\'\i}sica, Faculdade de Ci{\^e}ncias da Universidade de 
Lisboa, P-1649-003 Lisboa Codex, Portugal}
\affiliation{Kavli Institute for Theoretical Physics, Kohn Hall, University of California, Santa Barbara, CA 93106-4030}
\author{Ana Nunes}
\affiliation{Centro de F{\'\i}sica da Mat\'eria Condensada and 
Departamento de F{\'\i}sica, Faculdade de Ci{\^e}ncias da Universidade de 
Lisboa, P-1649-003 Lisboa Codex, Portugal}
\begin{abstract}
The dynamics of strongly immunizing childhood infections is still not well understood. Although reports of successful modeling of several incidence data records can be found in the literature, the key determinants of the observed temporal patterns have not been clearly identified. In particular, different models of immunity waning 
and degree of protection applied to disease and vaccine induced immunity have been 
debated in the literature on pertussis. Here we study the effect of disease acquired
immunity on the long term patterns of pertussis prevalence. We compare  five minimal models, all of which are stochastic, seasonally forced, well-mixed models of infection based on susceptible-infective-recovered dynamics in a closed population. These models reflect different assumptions about the immune response of naive hosts, namely total permanent immunity, immunity waning, immunity waning together with immunity boosting, reinfection of recovered, and repeat infection after partial immunity waning.
The power spectra of the output prevalence time series characterize the long term dynamics of the models.  For epidemiological parameters consistent with published data for pertussis, the power spectra show quantitative and even qualitative differences that can be used to test their assumptions by comparison with ensembles of several decades long
pre-vaccination data records. We illustrate this strategy on two publicly available
historical data sets.


\end{abstract}

\maketitle

\section{introduction}

Childhood infections remain a public health as well as a modeling challenge, despite many 
decades of control measures and theoretical efforts. Large vaccination programmes against some 
of these diseases started in the 1940s-1960s in the developed countries and led to dramatic 
reductions of incidence levels, but in the developing countries they are still a cause of 
significant levels of infant mortality \cite{trottier,bolkerweb,YorkeLondonII,Wallinga2001,broutin,Broutin2005b}. The resurgence of pertussis, also known as whooping cough, reported 
in developed countries after decades of high vaccine coverage \cite{CDC,Skowronski,deMelker,Crowcroft,pnaspertussis}, as well as a recent upsurge of measles in eastern and 
southern Africa  prompted a renewed interest in assessing the efficacy of control policies 
for these childhood infections.  

Such an assessment must rely on simulations based on mathematical models. The complexity and 
diversity of the long term dynamics of childhood diseases has been long acknowledged as a major 
problem in mathematical epidemiology \cite{Soper,YorkeLondonI,FineClarckson82,FineClarckson86}. More recent work focused on contact structure, stochasticity and 
seasonality has brought considerable advances in understanding and selecting some of the 
fundamental ingredients that drive the observed incidence temporal patterns \cite{Grenfnat2001,Grenfsci2000,b&earn,GrenfpD2001,wearing2005,MSimoes05062008,mixingpatterns,Read} and in developing analytic tools to deal with these ingredients 
and their interplay with the model's nonlinearities \cite{hethcote,vddrische,keelingrev,mckane1,PhysRevE.79.041922,VolzMeyers,andrea,forcedmckane,seasonal}.  

Models of higher complexity involve many parameters, some of which are difficult to determine, leaving  
considerable room for data fitting \cite{Metcalf,GrasslyFraser,p73,realnets}. On the other hand, 
one of the features of childhood infections is 
the variability exhibited by different data records, even within those that correspond to a single 
disease in comparable social environments \cite{broutin,b&earn,broutin2010}. 
Therefore, successful modeling of particular sets of incidence data records reported in the 
literature \cite{b&earn,rohaniinterplay,rohanistaged} has not closed the discussion 
about the key determinants of the observed dynamics. Measles is an exception for which a parsimonious
model has been shown to reproduce the behavior of different data sets \cite{Grenfnat2001,meslaesafrica,measlespre}, and it is generally accepted that the disease dynamics in 
large urban populations is adequately described by a stochastic seasonally forced well-mixed 
susceptible-infective-recovered (SIR) or susceptible-exposed-infective-recovered based model. 

At the opposite extreme, pertussis  
keeps defying mathematical modeling, as illustrated by recent efforts in different directions 
\cite{pnaspertussis,rohanistaged,rohanisignature,mckanewcough,rohanicontact,aguas}. In particular, several proposals explore
 hypothesis about disease induced and vaccine induced immunity \cite{pnaspertussis,broutin2010,aguas},
relying on assumptions about vaccine uptake and efficacy for the purpose 
of comparison with real data. This is a difficult problem of 
great interest and enormous relevance for public health. However, we are still lacking a sound 
uncontroverted model for pre-vaccination pertussis.
 
Here we have sought to contribute to the goal of using pre-vaccination data records
to obtain information about the properties of disease induced immunity.
The paper focuses on exploring the influence of naive hosts immune 
response in the long term patterns of pertussis prevalence as given by the 
averaged power spectra of simulated time series corresponding to several decades. 
The power spectrum of the stochastic seasonally 
forced well-mixed SIR model is compared with those of four modifications of the 
model that reflect different assumptions about disease induced immunity.  
These four different assumptions have been proposed in the literature in the
framework of deterministic models that have all been shown to be compatible with
available data.  The five variants  we consider 
deliberately avoid all the complications related with contact structure and spatial spread, 
as well as vaccine coverage and efficacy and waning of vaccine-induced immunity, in order to keep
a relatively low number of free parameters in the model. 

We show that the stochastic versions of the 
SIR model and its four variants have significantly different properties,
which translate into quantitatively  different prevalence and incidence power spectra.
This opens the possibility of using the stochastic properties of long, well resolved data records
to constrain these and other variants of the model, using  
the power spectra of the pertussis incidence time series in large urban centers in the pre-vaccination 
era as the target long term dynamics that the model should reproduce.
We illustrate this strategy by applying it to two publicly available 
historical data records for pre-vaccination pertussis incidence.

\section{models and methods}

\subsection{Models}

\begin{figure}
\includegraphics[trim=0.5cm 7.5cm 2.4cm 2.7cm, clip=true, width=0.6\textwidth]{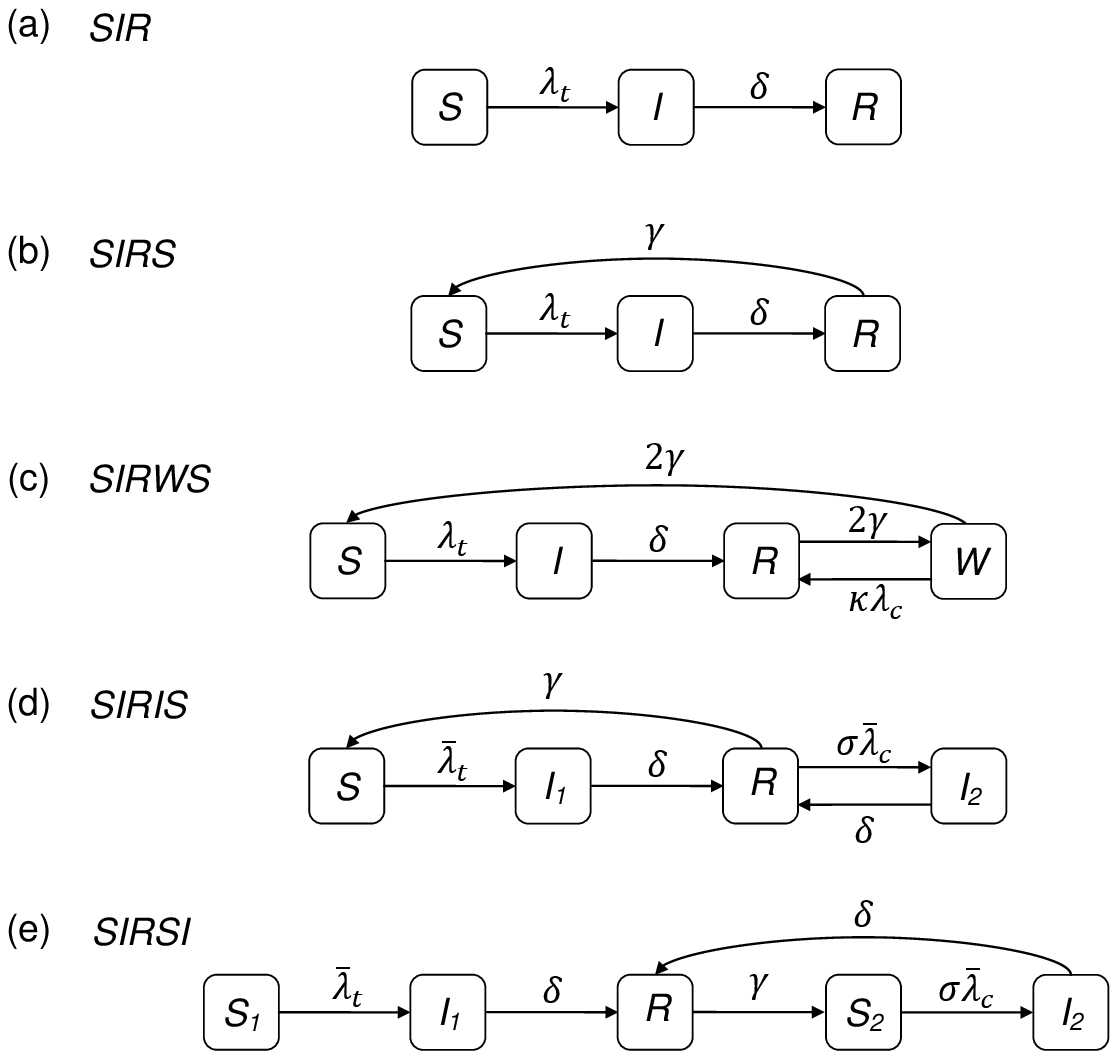}
\caption{The diagrams of five compartmental models for pertussis. The models (a) and (b) are the seasonally forced SIR and SIRS models, respectively, while the models (c), (d) and (e) are extensions of the SIR(S) paradigm. Model (c) allows for boosting of immunity proportional to, and potentially greater than, the force of infection, while model (d) allows for reinfection of recovered individuals. Model (e) accounts for a loss of infection-derived immunity and subsequent reinfection. In all diagrams, the transitions corresponding to the birth and death of individuals are not shown for simplicity. The meaning of the symbols associated with each transition is defined in the text.}
\label{diag}
\end{figure}

We consider five seasonally forced stochastic compartmental models summarized in Fig. \ref{diag} as 
candidates for the description of pre-vaccination pertussis (the deterministic counterparts of these models are described in the Supplementary Material). In all cases, the population includes three classes of individuals, the susceptible [$S$ in (a)-(d), $S_1$ and $S_2$ in (e)], the infectious [$I$ in (a)-(c), $I_1$ and $I_2$ in (d)-(e)] and the recovered [$R$ in (a)-(b) and (d)-(e), $R$ and $W$ in (c)]. Also in all five cases, there is replenishment through births of susceptible individuals that have never before contracted the disease. The birth rate is kept constant 
and equals the death rate $\mu$.

\subsubsection{SIR model}
Model (a) is the standard SIR model: recovery confers permanent immunity, and all infections 
are first infections. 
Infected individuals recover at a constant rate $\delta $, and susceptible individuals are infected 
at a rate $\lambda_t = \beta (t) I/N$, where $N$ is the population size, $I$ is the number of infected individuals and $\beta (t)$ is a periodic 
function of period one year that represents the variable contact rate associated with the school terms. 
For the purpose of  studying the long term temporal patterns of the fluctuations of this stochastic 
process the particular form of $\beta (t)$ is not crucial, and we shall take $\beta (t)=\beta_0 (1+\beta_1 \cos 2\pi t)$.

\subsubsection{SIRS model}
Model (b) is the standard SIRS model. It differs from the SIR model in that recovery does not confer 
permanent immunity. Instead, recovered individuals' immunity wanes at a constant rate $\gamma$ and 
then they become susceptible again. The fraction of 
primary infections is given approximately by $\mu /[\gamma(1-s^*-i^*) + \mu]$, where $s^*$ and $i^*$ are 
the equilibrium values for the susceptibles and infectives in the deterministic counterpart of the SIRS 
model (see the Supplementary Material for details on the deterministic SIRS model and on this calculation). Although there is no 
direct correspondence between primary infections and age groups in the model, it is reasonable to admit 
that a fraction of the infectives close to the fraction of non-primary infections, given by $\gamma(1-s^*-i^*)/[\gamma(1-s^*-i^*)+\mu]$, are not school children, and that 
$\beta_1$ should be reduced proportionally in this version of the model. This correction could be 
refined, but as we shall see in Section \ref{results} the properties of the stochastic fluctuations in this model are largely insensitive to the value of $\beta_1$.

\subsubsection{SIRWS model}
Model (c) is an SIRS-type model allowing for immune boosting of recovered individuals upon reexposure to infectious individuals. In this model, infectives $I$ recover at rate $\delta$ and susceptibles $S$ get infected at rate $\lambda_t$ as in the SIR model. However, in addition to the $S$ and $I$ classes, there are two classes of recovered individuals denoted by $R$ and $W$. The immunity of the former is 
not permanent and they move to the waning class $W$ at a constant rate $2\gamma$. The individuals in class $W$ undergo two possible transitions: further immunity loss at rate $2\gamma$ to become susceptible $S$, or immunity boosting upon contact with infectious individuals to return to the recovered class $R$ at a rate $\kappa \lambda_c=\kappa \beta_0 I/N$, where $\kappa$ is the immunity boosting coefficient. We call this scheme the immune boosting model and denote it by SIRWS.

An age-structured version of this model with vaccination has been used to explain the recent reemergence  of pertussis cases despite high vaccine coverage in Massachusetts, and also the shifts in total and age-specific incidences before and after mass vaccination \cite{pnaspertussis}.

\subsubsection{SIRIS model}
Model (d) proposed in \cite{aguas} sets a scenario based on the SIRS model with a moderate 
rate $\gamma$ of immunity loss where recovered individuals are immune to severe disease but
susceptible with reduced susceptibility to mild forms of the disease. The classes of individuals 
infected with severe and mild infections are denoted by $I_1$ and $I_2$, respectively.
Recovery from both forms of the disease takes place at the same rate $\delta$. Infectiousness of mild infections is reduced by a factor $\eta \in [0,1]$ and susceptibility of recovered individuals to mild infections is also reduced by a factor $\sigma \in [0,1]$ with respect to the susceptible. Moreover, since mild infections typically occur in adults that
are not affected by seasonal forcing the force of infection is therefore taken to be
$\bar {\lambda }_t  = (\beta(t) I_1 + \beta _0\eta I_2)/N$ for susceptible individuals $S$ and $\bar {\lambda }_c  = \beta_0 (I_1 + \eta I_2)/N$ for recovered individuals $R$. We call this scheme the 
reinfection model and denote it by SIRIS. 

The main feature of this model is that it exhibits a reinfection threshold \cite{gabrielareinfection}, that is, a value of infectiousness above which total disease incidence rises by one or more orders of magnitude, due to high incidence levels of mild infections. These mild infections are sub-clinical and so the disease incidence/prevalence and density of infectives in equilibrium for the deterministic counterpart of the SIRIS model is associated with the class $I_1$.

\subsubsection{SIRSI model}
Finally, model (e) proposed in \cite{rohanisignature} assumes an immune response that is a combination of (b) and (d), in the sense that recovered individuals are fully immune to the disease, but they lose this immunity at a certain rate $\gamma$ to become susceptible to repeat infections, although with reduced susceptibility. The two classes of susceptible individuals are denoted by $S_1$ for the naive susceptibles and $S_2$ for the susceptibles generated by immunity waning, and $\sigma \in [0,1]$ is the reduced susceptibility factor for repeat infections. The classes of individuals infected with first and repeat infections are denoted by $I_1$ and $I_2$, respectively.
Recovery from both classes takes place at the same rate $\delta $. The class of repeatedly infected individuals contributes with reduced infectiousness to the pool of infectives responsible for disease transmission, and $\eta \in [0,1]$ is the reduced infectiousness factor. Repeat infections play a similar role in this model to that of mild infections in the SIRIS model, and the force of infection is taken to be
$\bar {\lambda }_t  = (\beta(t) I_1 + \beta _0\eta I_2)/N$ for $S_1$ and $\bar {\lambda }_c  = \beta_0( I_1 + \eta I_2)/N$ for $S_2$, because repeat infections typically occur in adults whose contacts are not subject to school term forcing. Similarly to the SIRIS model, the disease incidence/prevalence and density of infectives in equilibrium for the deterministic counterpart of the SIRSI model is identified with the class $I_1$. We call this scheme the repeat infection model and denote it by SIRSI.

\begin{figure}
\includegraphics[trim=0cm 0cm 0cm 0cm, clip=true, width=0.48\textwidth]{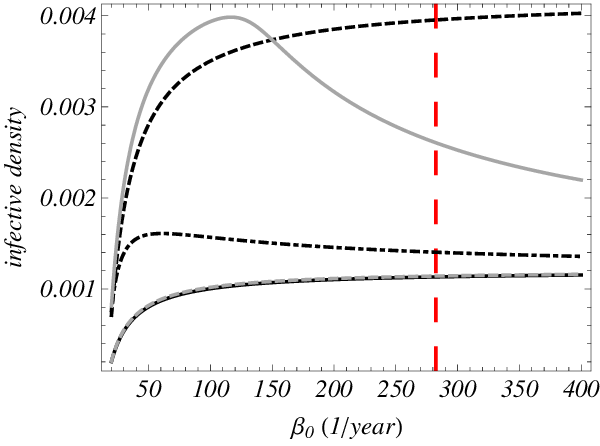}
\caption{(Color online) Infective density in the equilibrium of the unforced deterministic versions of the models, as a function of $\beta_0$. The line stands for the $I/N$ in the SIR (solid black), SIRS (dashed black) and SIRWS (dot-dashed black), and for the $I_1/N$ in the SIRIS (solid gray) and SIRSI (dashed gray) model. The range of $\beta_0=R_0(\delta+\mu)$ in the plot corresponds to $R_0 \in [1.2,24]$, $\mu=1/50$ $\text{year}^{-1}$ and $\delta=365/22$ $\text{year}^{-1}$. Here $R_0$ is the so-called basic reproductive ratio of a disease defined as the average number of secondary cases generated by one infected individual in a fully susceptible population during one infectious period \cite{amay}. The red long-dashed vertical line indicates the value of $\beta_0$ for which $R_0=17$. Parameters: $1/\gamma=20$ years (SIRS model); $1/\gamma=10$ years (SIRWS model, \cite{pnaspertussis}); $1/\gamma=20$ years, $\eta=0.5$, and $\sigma=0.25$ (SIRIS model, \cite{aguas}); $1/\gamma=20$ years, $\eta=0.1$, and $\sigma=0.3$ (SIRSI model).
}
\label{fig8}
\end{figure}

We finish the description of the models by pointing out that, for a fixed $\beta_0$, the unforced ($\beta_1=0$) deterministic counterparts of the five stochastic models introduced in this section have different densities of infectives in equilibrium. In Figure \ref{fig8} we plot these equilibrium densities as a function of $\beta_0$. The values of the remaining parameters will be justified later.  
For $0<\beta_1\leq 0.1$, the density of infectives in the deterministic versions of the five models oscillates with the period 1 year around the equilibrium value of the corresponding unforced equations (see the Supplementary Material).
Note that the unforced SIRWS model may have also limit cycles as stable attractors for some parameter values. This is, however, not the case for the parameter values that we will use in our analysis (throughout the main text $\beta_0$ is fixed at the dashed vertical line in Fig. \ref{fig8}).

\subsection{Data records}

\begin{figure}
\includegraphics[trim=0cm 0cm 0cm 0cm, clip=true, width=0.8\textwidth]{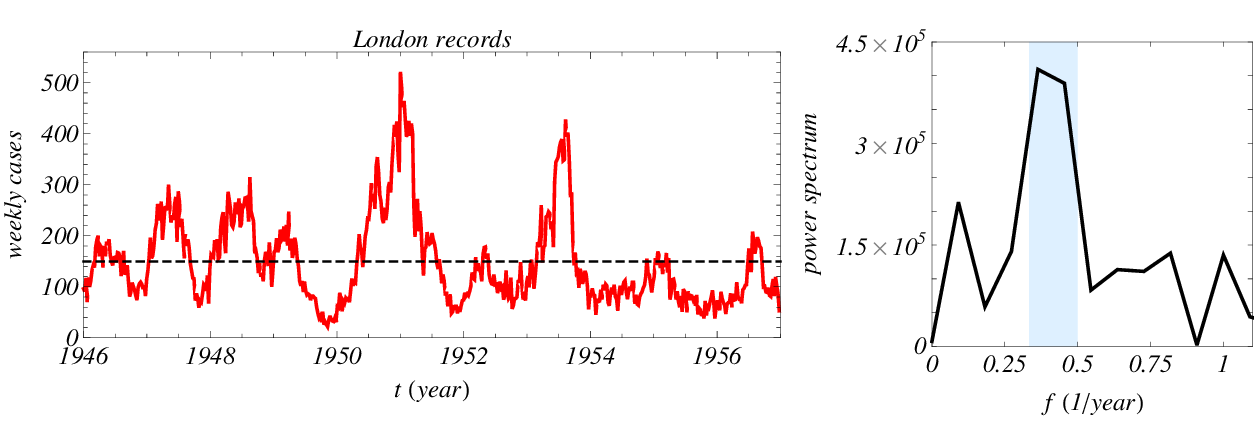}
\caption{(Color online) Left: new weekly cases of pertussis in (Greater) London before vaccination. The dashed horizontal line is the average number of weekly cases. Right: spectrum of the time series.}
\label{fig5}
\end{figure}

\begin{figure}
\includegraphics[trim=0cm 0cm 0cm 0cm, clip=true, width=0.8\textwidth]{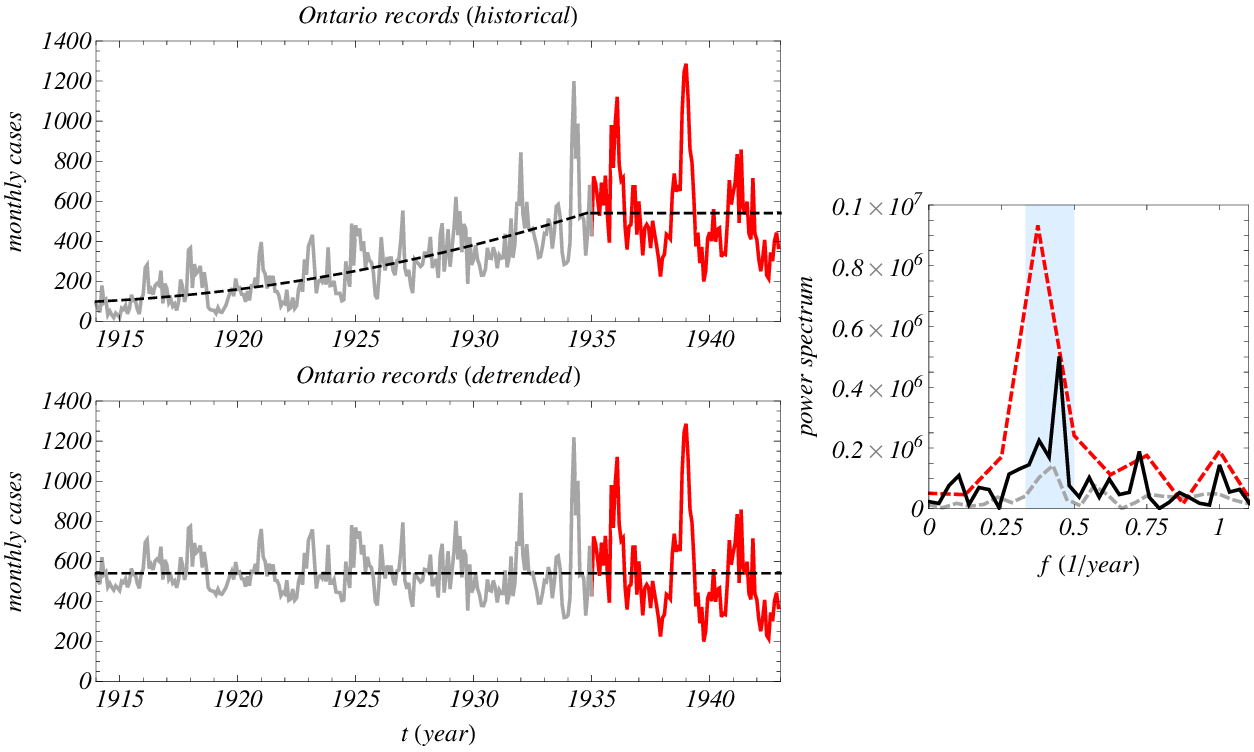}
\caption{(Color online) Left: new monthly cases of pertussis in the Canadian province Ontario before vaccination (upper panel) and the detrended time series (lower panel). The dashed horizontal line in the lower panel is the average number of monthly cases. Right: the solid black line is the spectrum of the full time series shown in the lower left panel, the red (black) dashed and the gray dashed lines are the spectra calculated from the detrended and undetrended partial time series shown in the same color.} 
\label{fig6}
\end{figure}

In this paper, we consider the historical data records for pre-vaccination pertussis incidence in Greater London in the period 1946-1957 (Figure \ref{fig5}, the left panel) and in Ontario in the period 1914-1943 (Figure \ref{fig6}, the left panels). The time series correspond to weekly data, in the case of London, and monthly data in the case of Ontario. Both raw and detrended (but not rescaled) data are shown for Ontario, where significant changes in population size and reporting efficiency are apparent during most of that period. In order to analyze the populations involved in these two data records, we shall consider only the undetrended data for Ontario corresponding to the last 8 years of the 29 years period. The ratio of the populations is close to 8/3.5, according to demographic data that set the population of Greater London close to 8 million and that of Ontario close to 3.5 million in the period under consideration. However, the average recorded rate of new cases in a month interval, which is similar for the two data sets, points to a ratio of effective populations close to 1. Differences in surveillance coverage and/or in reporting efficiency could explain this. For London, levels of pertussis reporting efficiency in the range 5\%-25\% have been acknowledged in the literature \cite{reportpertussis}. The fact that the effective population for Ontario is approximately of the same size indicates that a higher reporting efficiency compensates for the smaller population. 

It is known from the case notification data from England and Wales and other locations that the inter-epidemic periods for pertussis in the pre-vaccination era show significant multiannual structure. The multiannual periods lie in the range of 2-3 years and additionally annual outbreaks can be observed \cite{b&earn,rohanisignature,synchrony}. The power spectra computed the data confirm these observations (see the right panels in Fig.\ref{fig5} and in Fig. \ref{fig6}). In the spectra plots, the shaded region marks the range of frequencies, $f \in [1/3,1/2]$ $\text{year}^{-1}$, corresponding to the interepidemic periods $1/f\in [2,3]$ years. Because of limited length and resolution in time, the spectra have a short range and a low resolution in frequency, however this affects mostly the annual peak. The widths of the multiannual peaks are comparable with each other. For Ontario, the power spectra of the full time series and of the detrended and undetrended partial data sets show the same dominant frequency components. In what follows, we shall consider only the power spectrum of the undetrended data.

\subsection{Parameter values}

\begin{table}[t]
\centering
{\begin{tabular}{ccc}
\hline\hline Epidemiological meaning & Notation & Value \\ 
\hline \textit{Per capita} birth/death rate &  $\mu$  &  $1/50$ $\text{year}^{-1}$  \\
\hline Average lifespan &  $1/\mu$  & $50$ years \\ 
\hline Rate of recovery from infection &  $\delta$   & $365/22$ $\text{year}^{-1}$  \\
\hline Average infectious period &  $1/\delta$  &  $22$ days \\     
\hline Basic reproductive ratio &  $R_0$  &  $17^{\dagger}$  \\
\hline Average contact rate  &  $\beta_0$   &  $282.39$ $\text{year}^{-1}$ \\
\hline Amplitude of seasonal forcing &   $\beta_1$  &  $[0,0.1]$ $\text{year}^{-1}$  \\
\hline Rate of loss of naturally acquired immunity  &  $\gamma$  &  $[1/40,1/10]$ $\text{year}^{-1}$ \\ 
\hline Average infection-derived immunity period & $1/\gamma$  &  $[10,40]$ years  \\ 
\hline Relative infectiousness of repeat to primary (SIRSI)&  $\eta$  & $[0,1]$ \\
or mild to severe infections (SIRIS)&  &  \\
\hline Relative susceptibility of repeat to primary (SIRSI)&  $\sigma$  & $[0,1]$ \\
or mild to severe infections (SIRIS)&  &  \\
\hline Boosting coefficient & $\kappa$ & $[0.2,20]$ \\
\hline\hline
\end{tabular}
\caption{The epidemiological description of the parameters and the range of their values for pertussis. $^{\dagger}$ For the SIRS and SIRSI models, the
whole range of possible $R_0$ values estimated from the average age at first infection  is 
explored in the Supplementary Material.}
\label{table1}}
\end{table}

The values of the demographic and epidemiological parameters that are well established in independent 
data sources are kept fixed (see, for example, \cite{b&earn,amay}). These are the birth/death rate $\mu$ (or the average lifespan $1/\mu$), the rate of recovery from infection $\delta$ (or average infectious period $1/\delta$) and the average contact rate $\beta_0$. The value of the latter corresponds to the value of the basic reproductive ratio $R_0=\beta_0/(\delta+\mu)$ reported in the literature for the two data sets considered in this study \cite{b&earn}. If the reported value of $R_0$ comes from average age at infection data, together with the assumption of SIR dynamics, then it has to be corrected for models with temporary immunity. In the Supplementary Material we explore the whole range of possible $R_0$ values for the SIRS and SIRSI models and show that the analysis of Section \ref{results} below is not affected by this correction. We also give for the SIRWS and SIRIS models the dependence of $R_0$ on the remaining model parameters,  computed from average age at infection data on the assumption that the naive susceptibles are homogeneously mixed in the whole population.

For the remaining parameters we either use the estimates found in the literature for a particular model or explore an appropriate range of possible values if such estimates are absent. For example, the accepted range for the duration of naturally acquired immunity for pertussis is 7 to 20 years \cite{review_im_pertussis}. For the SIRS model, we take $1/\gamma$ equal 20 to 40 years. On one hand, the large upper limit allows for a comparison with the SIR model. On the other hand, as we shall see, taking $1/\gamma$ smaller than 20 years would make the prediction of the model worse. For the SIRIS model, we take $1/\gamma=20$ years \cite{aguas} and the relative infectiousness and relative susceptibility of mild infections $\eta,\sigma \in [0,1]$. The prediction of this model for other values of $1/\gamma$ can still be done using the SIRS model as one of the limiting cases of the SIRIS model. For the SIRWS model, we take the value of the boosting coefficient $\kappa=20$ and $1/\gamma=10$ years considered in Ref. \cite{pnaspertussis} and in addition we study the behavior of the model for $1/\gamma=40$ years and $\kappa \in [0.2,10]$. For the SIRSI model, we set $1/\gamma=20$ years and the relative infectiousness and relative susceptibility of repeat to primary infections $\eta,\sigma \in [0,1]$ \cite{rohanisignature}. We discuss in the Supplementary Material the behavior of this model for other values of the duration of immunity.

For practical purposes we use only three values of the amplitude of seasonal forcing $\beta_1=0,0.05,0.1$. We will however discuss the predictions of the models for intermediate values of $\beta_1$ whenever necessary. The summary of the parameter values is given in Table \ref{table1}.

\subsection{Stochastic simulations}
To study the behavior of the five candidates for the description of pre-vaccination pertussis we use stochastic simulations of the processes described in Fig. \ref{diag}. The simulations are based on a modification of Gillespie's algorithm \cite{Gillespie} which accounts for the explicit time dependence in the contact rate \cite{DavidAnderson},\cite{Lu}. In Sections \ref{sir}-\ref{sirsi}, each simulation run starts from a random initial condition and the prevalence of the disease is recorded with a time step of 0.05 year for 450 years after 50 years of transient. From each simulation run the power spectrum of the prevalence time series is computed with the use of the discrete Fourier transform. The final spectra are averages of $10^3$ simulations. The population size, $N$, used in the simulations in Sections \ref{sir}-\ref{sirsi} is $5\times10^5$.

\section{results}
\label{results}

We investigate the dynamics of the stochastic seasonally forced SIR, SIRS, SIRWS, SIRIS and SIRSI models by comparing the power spectra of long time series of the five models in the relevant parameter ranges. We first describe the performance of the SIR model and then assess whether each of its four extensions improves or worsens the results for the SIR model. In all cases, the spectra correspond to the fluctuations around the equilibrium of the unforced deterministic versions of each model ($\beta_1=0$) or around the associated period one year stable attractor ($\beta_1>0$). Numerical and simulational results for the parameter ranges we have explored never showed evidence of other attractors. A direct quantitative comparison of these power spectra with the spectra shown in Fig. \ref{fig5} and Fig. \ref{fig6} based on the amplitude of the multiannual peaks and their precise location is undermined by the low resolution in frequency and poor statistics of the two data sets, which correspond to much shorter periods than the ones that can be used in the numerical simulations. That is why our first criterion in making the comparison between the models' predictions and the spectra computed from the data will be the position of the multiannual peak. In order to be able to explain historical data records of pertussis incidence a model's spectrum should have the multiannual peak located in the range of frequencies, $f \in [1/3,1/2]$ $\text{year}^{-1}$ (see Fig. \ref{fig5} and Fig. \ref{fig6}). This region is the shaded region shown in all spectra plots throughout the paper. 

As pointed out before, in Sections \ref{sir}-\ref{sirsi} we use a typical population size $N=5\times10^5$ in all models and compute spectra from the prevalence time series. Changing population size and/or computing the incidence in a given interval instead of the prevalence of the disease may change somewhat the amplitude of the multiannual and annual peaks but does not influence their position.

\begin{figure}
\includegraphics[trim=0cm 0cm 0cm 0cm, clip=true, width=0.2666\textwidth]{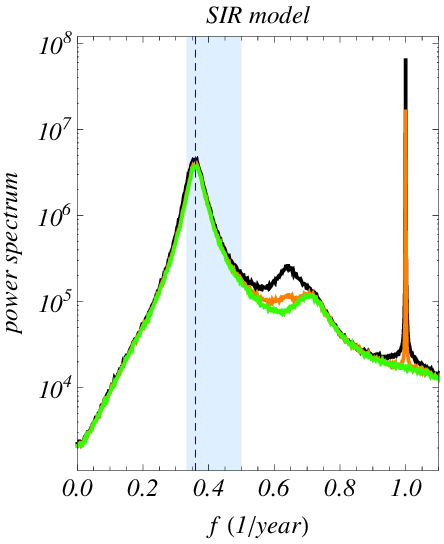}
\caption{(Color online) The average power spectra from simulations of the SIR model for different amplitudes of seasonal forcing, $\beta_1$. Parameters: $\beta_1=0$ [green (light gray)], $0.05$ [orange (dark gray)], $0.1$ (black). The multiannual peaks of the spectra computed from the London and Ontario data sets are located in the shaded region. Henceforth, all simulation spectrum plots will be shown in lin-log scale and the dashed line will indicate the dominant frequency of the stochastic multiannual peak found from the simulations.}
\label{fig0}
\end{figure}
 
\subsection{SIR model}
\label{sir}

The model's power spectrum as a function of frequency for different values of the forcing amplitude $\beta_1$ is shown in Fig. \ref{fig0}. The power spectra computed from the stochastic simulations exhibit two types of peaks: a well defined dominant annual peak due to the deterministic annual limit cycle and a subdominant broad multiannual stochastic peak with the shape and the main frequency similar to that of the unforced case. The increase in $\beta_1$ results in a more enhanced annual peak which means that the contribution of annual epidemics in the time series increases as $\beta_1$ increases. The multiannual stochastic peak reflects the presence of noisy oscillations in the incidence time series with that dominant frequency. The mechanisms by which internal noise excites these resonant fluctuations has been treated in several papers \cite{mckane1,PhysRevE.79.041922,forcedmckane,seasonal,mckanewcough}. They can be described analytically using van Kampen's expansion of the master equation of the underlying stochastic process around the attractor of the deterministic system \cite{vankampen}. We point out that in this study the dominant frequency of the main stochastic peak in the seasonally forced models can be computed from the unforced model whose analysis is easier (see Supplementary Material for more information on the analytical computation of the power spectra). 

For the SIR model, the frequency and the shape of the dominant stochastic peak are largely independent of $\beta_1$, and the peak's frequency lies in the shaded region (see the Supplementary Material for the analytical expression of the spectrum). The dominant period of stochastic fluctuations in the SIR model is 2.7 years. In Sections \ref{sirs}-\ref{sirsi} we will discuss whether the SIRS, SIRWS, SIRIS and SIRSI extensions of the SIR model improve or worsen its performance. In particular, we will be interested to know how the position and the amplitude of the stochastic peak change with respect to that of the SIR model and the robustness of the behavior of each model under the variation of free parameters.

\subsection{SIRS model}
\label{sirs}
\begin{figure}
\includegraphics[trim=0cm 0cm 0cm 0cm, clip=true, width=0.8\textwidth]{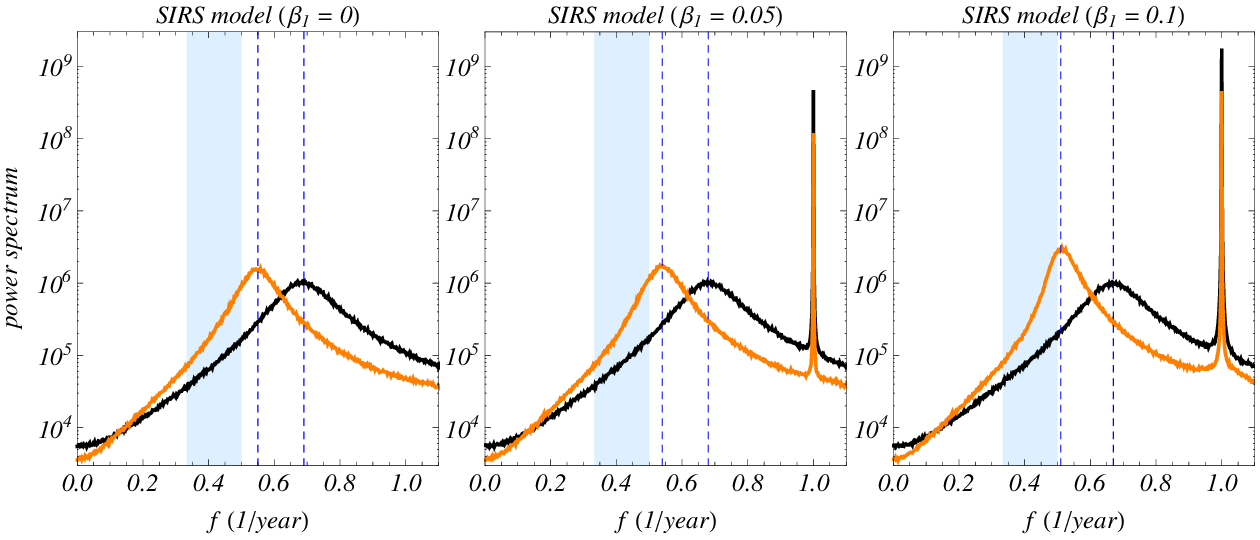}
\caption{(Color online) The average power spectra from simulations of the SIRS model for different amplitudes of seasonal forcing, $\beta_1$, and two values of the immunity waning rate, $\gamma$. Parameters: $1/\gamma=20$ years (black), $40$ years [orange (gray)]. The multiannual peaks of the spectra computed from the London and Ontario data sets are located in the shaded region. The dashed lines indicate the dominant frequencies of the stochastic multiannual peaks found from the simulations.}
\label{fig1}
\end{figure}

In the SIRS model, the rate of immunity waning $\gamma$ is a free parameter. In the case of the lifelong immunity, $\gamma=0$, the SIRS model reduces to the SIR model considered in the previous section. Figure \ref{fig1} shows the model's spectrum for different values of the forcing amplitude $\beta_1$ and two typical values of $\gamma$. In the unforced SIRS model, $\beta_1=0$, the spectrum has a pronounced stochastic peak (see the Supplementary Material for the analytical expression of the spectrum). However, it is situated outside the region of interest both for $1/\gamma=40$ years and for $1/\gamma=20$ years. 
For fixed $\gamma$ and increasing $\beta_1$, the frequency of the stochastic peak is slightly shifted towards the target region but it remains outside it even for large values of $\gamma$ and, at the same time, the power associated with the annual peak becomes very large. Note that if $\beta_1$ is decreased to take into account that non primary infections should not be subject to seasonal forcing in this model the dominant frequency of the stochastic peak is moved even further away from the shaded region. We conclude by noting that the SIRS model's spectrum is incompatible with the London and Ontario data unless in the limit of $\gamma\to0$ when it approaches the SIR model. The performance of the SIRS model for other values of $R_0$ is also discussed in the Supplementary Material.

\begin{figure}
\includegraphics[trim=0cm 0cm 0cm 0cm, clip=true, width=0.8\textwidth]{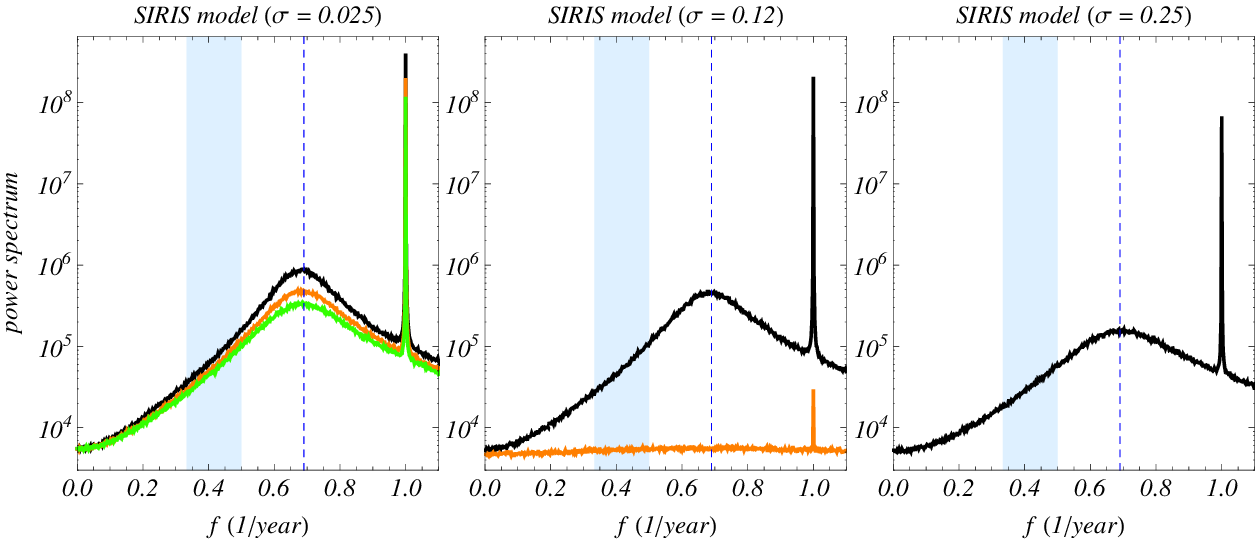}
\caption{(Color online) The average power spectra from simulations of the SIRIS model for $\beta_1=0.05$, $1/\gamma=20$ years and different values of the relative infectiousness, $\eta$, and relative susceptibility, $\sigma$, of mild to severe infections. The spectra for other values of $\beta_1$ are given in the Supplementary Material. Parameters: $\eta=0.1$ (black), $0.5$ [orange (dark gray)] and $0.75$ [green (light gray)]. The values of $\eta$ and $\sigma$ were chosen from $(0,1)$ so as to guarantee that the equilibrium incidence in the unforced deterministic SIRIS model is of the same order of magnitude as in the SIR(S) models. The multiannual peaks of the spectra computed from the London and Ontario data sets are located in the shaded region. The dashed lines indicate the dominant frequencies of the multiannual peaks found from the simulations.}
\label{fig2}
\end{figure}

\subsection{SIRIS model}

Next we analyze the power spectra from simulations of the SIRIS model. This model has three free parameters apart from the amplitude of seasonal forcing $\beta_1$: the rate of immunity loss $\gamma$, and the relative infectiousness $\eta$ and relative susceptibility $\sigma$. For the limiting case of $\eta=0$ and $\sigma=0$ the model reduces to the SIRS model whose multiannual peaks are located outside the shaded region. The model's behavior is, however, well understood if we vary only $\eta$ and $\sigma$. The results are shown in Fig. \ref{fig2} for fixed $\beta_1=0.05$ and $1/\gamma=20$ years. The spectrum for small values of $\eta$ and $\sigma$ is reminiscent of the SIRS model. A comparison of the spectra in all panels of Fig. \ref{fig2} shows that the  stochastic and deterministic peaks are the highest for the smallest values of both $\eta$ and $\sigma$. This is because it is easier to generate incidence oscillations in this parameter range. The increase in the values of $\eta$ and $\sigma$ results in a flattening of the spectrum occurring through a gradual disappearance of, first, stochastic and, then, deterministic peaks. Moreover, the study of the dependence of the power spectrum on the amplitude of seasonal forcing $\beta_1$ does not change this picture. For example, for the smallest values of $\eta$ and $\sigma$ used in Fig. \ref{fig2} we observe that the position and the shape of the stochastic peaks agree perfectly for $\beta_1 \in [0,0.1]$ and that they are always outside of the region of interest (this result is given in the Supplementary Material). The parameter region with higher values of $\eta$ and $\sigma$ is even less interesting because of the flattening of the spectrum. Thus for all relevant ranges of $\eta$, $\sigma$ and $\beta_1$ the stochastic multiannual peak in this model is situated outside the region of interest except when it approaches the limit of the SIR model.

\begin{figure}
\includegraphics[trim=0cm 0cm 0cm 0cm, clip=true, width=0.8\textwidth]{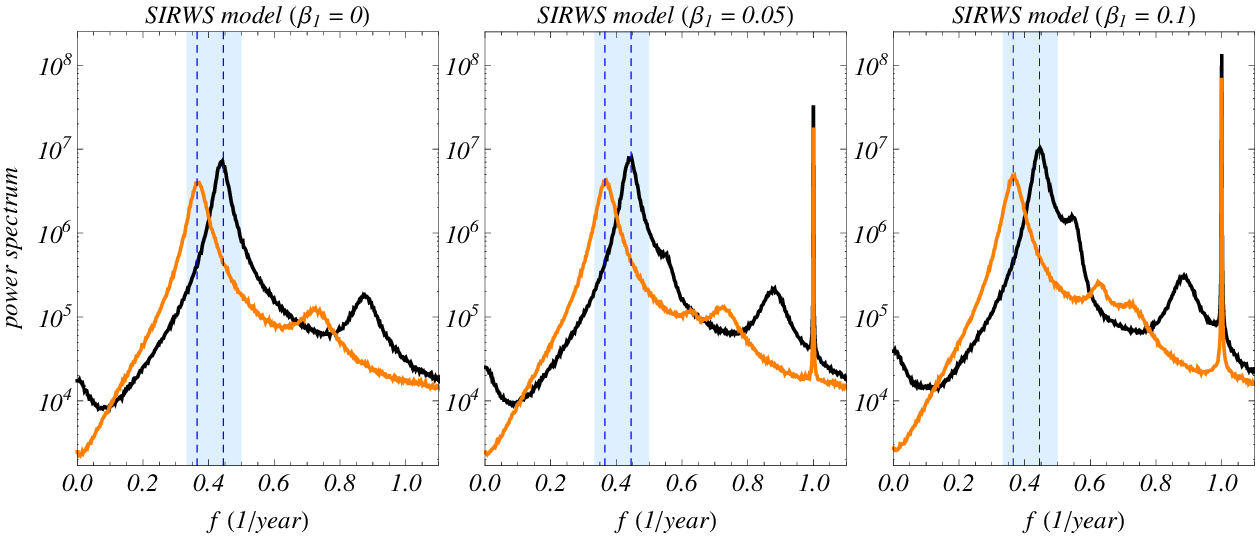}
\caption{(Color online) The average power spectra from simulations of the SIRWS model for different amplitudes of seasonal forcing, $\beta_1$, and two values of the immunity waning rate, $\gamma$. Parameters: $1/\gamma=10$ years (black) and $40$ years [orange (gray)]. The immune boosting coefficient $\kappa=20$ in all panels. The spectra for other values of $\kappa$ are given in the Supplementary Material. The multiannual peaks of the spectra computed from the London and Ontario data sets are located in the shaded region. The dashed lines indicate the dominant frequencies of the multiannual peaks found from the simulations.}
\label{figure9}
\end{figure}

\subsection{SIRWS model}

Here we show the power spectra computed from simulations of the SIRWS model for boosting coefficient $\kappa=20$ and different values of $\gamma$ and $\beta_1$, see Fig. \ref{figure9}. The results for $\kappa=0.2,1$ and 10 are given in the Supplementary Material and will be discussed later in this section. In the absence of seasonality, $\beta_1=0$, the spectra have a dominant multiannual peak situated in the target region for the whole range of $1/\gamma \in [20, 40]$ years. When $\beta_1$ increases a deterministic annual peak appears in the spectrum while the dominant frequency of the multiannual peak stays unchanged. Similarly to the SIR model, the higher the amplitude of seasonal forcing $\beta_1$ is the higher the annual peak is. However, unlike in that model, this phenomenon is accompanied by an overall complication of the structure of the spectrum for higher values of $\beta_1$. For example, the power spectrum for $1/\gamma=10$ years and $\beta_1=0.1$ (black line in the right panel of Fig. \ref{figure9}) has several secondary multiannual peaks, one of which is situated near but outside the shaded region. Thus, the SIRWS model predicts a rather broad range of frequencies corresponding to the stochastic fluctuations around the annual deterministic cycle.

Finally, we discuss the results for $k\neq20$ (see Supplementary Material for details). For $\kappa=0.2,1,10$ and $1/\gamma=40$ years, the stochastic multiannual peaks of all power spectra lie in the shaded region. The same happens for $\kappa=10$ and $1/\gamma=10$ years but not for $k=0.2,1$. For $\kappa\to 0$ and $\gamma\gg\mu$ the SIRWS model exhibits a SIRS-like behavior, thus the multiannual peaks of the SIRWS model are outside of the region of interest likewise in that model.

\subsection{SIRSI model}
\label{sirsi}

\begin{figure}
\includegraphics[trim=0cm 0cm 0cm 0cm, clip=true, width=0.8\textwidth]{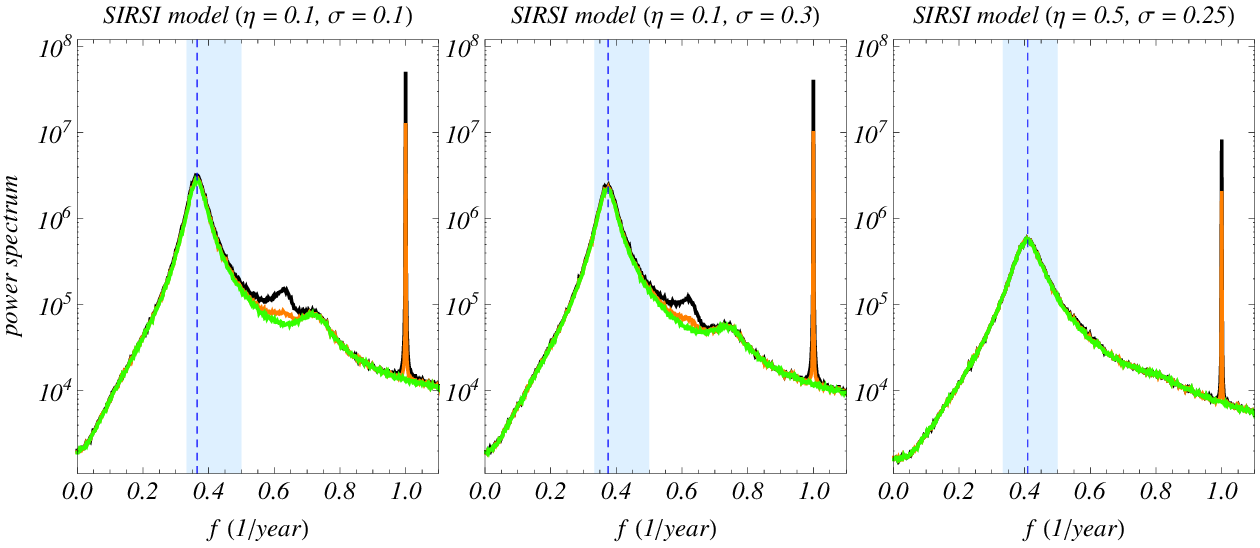}
\caption{(Color online) The power spectrum from simulations of the SIRSI model. Parameters: $\beta_1=0$ [green (light gray)], $0.05$ [orange (dark gray)] and $0.1$ (black), $1/\gamma=20$ years. The values of the relative infectiousness $\eta$ and the relative susceptibility $\sigma$ of repeat to primary infections were chosen from a set of all possible values explored in the Supplementary Material. The multiannual peaks of the spectra computed from the London and Ontario data sets are located in the shaded region. The dashed lines indicate the dominant frequencies of the multiannual peaks found from the simulations.}
\label{fig3}
\end{figure}

We finish by presenting the results from simulations of the SIRSI model. Figure \ref{fig3} shows the spectra for different values of $\beta_1 \in [0,0.1]$, $\sigma,\eta \in (0,1)$ and $1/\gamma=20$ years. Similarly to the SIR model, the simulation spectra of the SIRSI model have a high stochastic multiannual peak in addition to the annual peak due to the deterministic limit cycle. The amplitude of the deterministic peaks and therefore the power associated with them increase as $\beta_1$ increases while the stochastic peaks stay almost unchanged. The increase, however, is smaller in the SIRSI model than in the SIR model. This means that, notwithstanding the similarity of their simulation spectra, the contribution of annual epidemics is less in the former than in the latter, corresponding to a qualitative difference in the times series of the two models. 

For three sets of parameter values chosen in Fig. \ref{fig3} the dominant stochastic multiannual peaks lie in the shaded region, moreover their frequencies and shapes are largely independent of $\beta_1$. To explore the consequences of varying $\eta$ and $\sigma$ for fixed $\gamma$, we considered in the interval $(0,1)$ nine equally spaced values and constructed a grid of 81 points in the $(\eta,\sigma)$ space. As the position and the shape of the stochastic peak in this model is independent of $\beta_1$ we can calculate the stochastic peak's frequency from the unforced model (see Supplementary Material for more details). For example, for $1/\gamma=20$ years used in Fig. \ref{fig3}, the number of spectra for the SIRSI model with the stochastic peak within the shaded region is 62 (for $1/\gamma \in [10,40]$ years this number varies from 43 to 78). We thus conclude that the SIRSI model's spectrum illustrated in Fig. \ref{fig3} is robust with respect to variation of all free parameters, and that it exhibits some degree of variability in the amplitude of the stochastic multiannual peak as parameters are varied within their accepted ranges. The robustness of these results with respect also to variations of $R_0$ keeping the average age at first infection fixed at its value for the SIR model is illustrated in the Supplementary Material.

\subsection{Comparison with real data}

\begin{figure}
\includegraphics[trim=0cm 0cm 0cm 0cm, clip=true, width=0.8\textwidth]{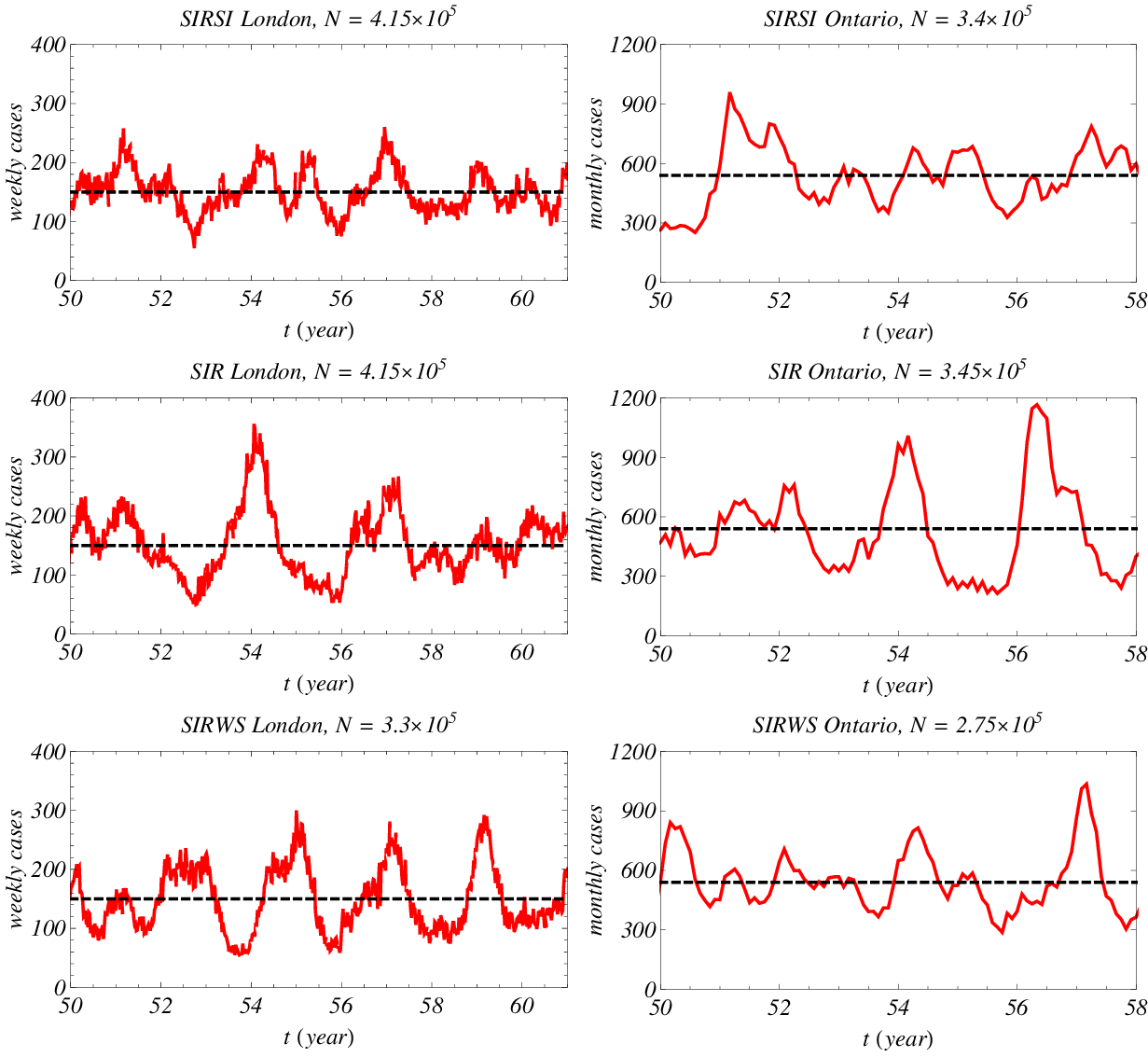}
\caption{(Color online) Incidence of the disease in a simulation of the SIRWS, SIR and SIRSI models. The horizontal dashed line is the average number of weekly (London) or monthly (Ontario) cases (compare these with the dashed lines in Fig. \ref{fig5} and in Fig. \ref{fig6}). Parameters: $\beta_1=0.05$ (all plots); $\kappa=20$, $1/\gamma=10$ years (SIRWS model, the black line in the middle plot of Fig. \ref{figure9}); $1/\gamma=20$ years, $\eta=0.1$, $\sigma=0.1$ (SIRSI model Ontario) and $\sigma=0.3$ [SIRSI model London, the orange (dark gray) line in the middle plot of Fig. \ref{fig3}]. Note that the level of seasonal forcing $\beta_1$ does not change the average incidence.}
\label{figlast}
\end{figure}

The results for the power spectra of stochastic simulations show that out of the four extensions of the SIR model under consideration only 
SIRWS and SIRSI have frequency values for the stochastic multiannual peak that preserve or improve the compatibility with data for 
pre-vaccination pertussis previously obtained with the SIR model. Here we compare the other features of spectra of the SIR model and of the SIRWS and SIRSI extensions with the spectra for London and Ontario data sets shown in Figs. \ref{fig5} and \ref{fig6}. 

As a first step towards achieving this we determine the population size $N$ for each model. As discussed before both data sets have the similar average recorded rate of new cases in a month interval and thus correspond to a similar effective population. However, the deterministic counterparts of the models, and consequently the stochastic models too, predict different densities of infectives in the steady state. To resolve this, for each of the three models we make simulations with the same length and the same sampling time as those of the London or Ontario data sets and record the incidence of the disease in the sampling interval. The transient period in all simulations is taken to be 50 years and each simulation starts from a random initial condition. We calibrate the effective population size $N$ for a given model by imposing that the incidence averaged over $10^3$ different simulation runs is the same as the time averaged incidence in the corresponding data set, given by the dashed line in Fig. \ref{fig5} and in Fig. \ref{fig6}. In Fig. \ref{figlast} we show typical incidence time series for $\beta_1=0.05$ and the effective population size $N$ for each of the three models from which individual power spectra are computed. 

The second step is to determine for each model the level of seasonality  that corresponds to the data set. We estimate the value of the amplitude of the seasonal forcing $\beta_1$ as the value for which the amplitude of the annual deterministic peak in the power spectrum of incidence times series averaged over $10^3$ different simulation runs is equal to the amplitude of the annual peak in the spectrum computed from the data.  

\begin{figure}
\includegraphics[trim=0cm 0cm 0cm 0cm, clip=true, width=0.8\textwidth]{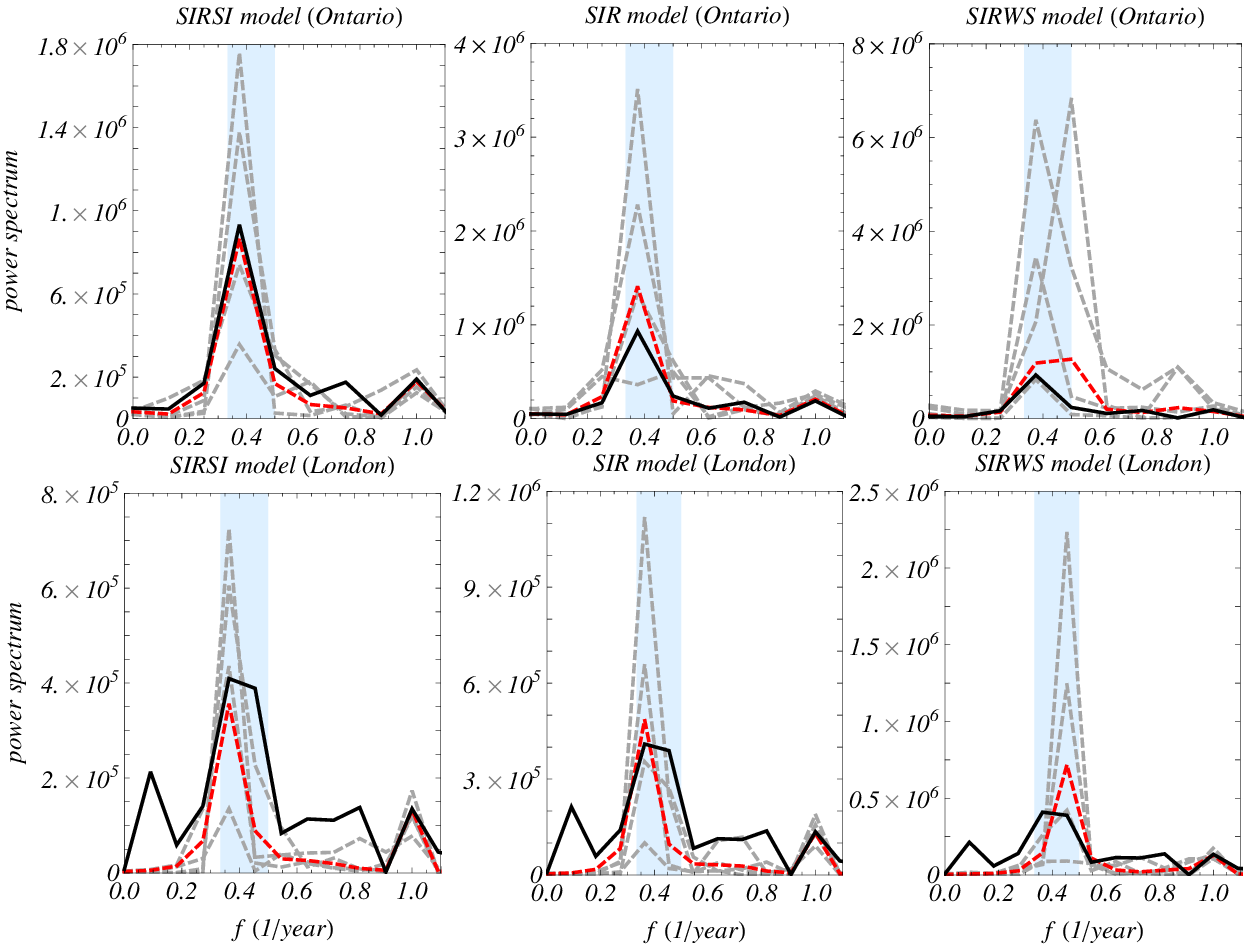}
\caption{(Color online) The solid black line is the power spectrum of the London or Ontario data sets as indicated in each panel. The dashed red (black) line is the average power spectrum of the incidence time series computed from $10^3$ simulation runs with the same length and the same sampling time as those of the London or Ontario data sets. The dashed gray lines are four typical power spectra computed from an individual simulation. Parameters: $\beta_1=0.06$ (SIRSI Ontario), $\beta_1=0.07$ (SIRSI London), $\beta_1=0.05$ (SIR Ontario), $\beta_1=0.06$ (SIR London), $\beta_1=0.03$ (SIRWS Ontario), $\beta_1=0.05$ (SIRWS London). The remaining parameter values are as in Fig. \ref{figlast}.}
\label{figconc}
\end{figure}

Using the values of $N$ and of $\beta_1$ determined in this way,
we produce from the stochastic models time series mimicking the London or Ontario data sets, and average their power spectra over $10^3$ simulation runs. The power spectra from simulation [dashed red (black) lines] and the spectra computed from the data sets [black solid lines] are shown in Fig. \ref{figconc}. The power spectra obtained with the SIR model have stochastic multiannual peaks with amplitudes that are similar, though on average larger, than those of the data sets power spectra.
For the SIRWS extension, the average amplitude of the stochastic peak is even larger, 
while the SIRSI extension gives a better agreement with this feature of the data with respect
to the SIR model.

The results of Fig. \ref{figconc} thus suggest that the SIRSI model is the one that 
best reproduces 
the stochastic properties of both data sets. However, as shown by the four examples plotted in the 
figure for each case (dashed gray lines), the poor statistics of the short term 
samplings translates into a considerable variability in the power spectra computed from each 
time series. The two historical data records considered here illustrate the principle that
the incidence power spectra can be used to discriminate the three variants of the model, 
but a final conclusion on their performance would require
the analysis of longer time series. 

\section{Discussion and Conclusions}
\label{lastsection}

Different assumptions about disease and vaccine induced immunity have been proposed in the recent
literature to model the dynamics of pertussis in the presence of mass vaccination. 
Our motivation has been to study separately the influence of disease acquired immunity only,
in order to reduce the number of free parameters, and to consider explicitly the stochastic 
properties of the incidence time series as part of the model's output, in order to look for 
further constraints.

We have used the basic SIR model and four variants that reflect different immune responses. We
characterized the dynamic patterns of each of these five models through the average power spectra
of an ensemble of long term prevalence time series obtained from stochastic simulations.
We have found that the power spectra of the five alternative models show quantitative 
and even qualitative differences. One of them (SIRS) fails to reproduce a stochastic peak
in the frequency range found in real data for pertussis, and another one (SIRIS) is too stiff
to produce stochastic fluctuations with the observed amplitudes. 

Assessing the performance of the other two variants, SIRWS and SIRSI, with respect to the SIR model
requires a quantitative comparison of the 
simulated power spectra with real data, for which purpose the effective population size and the
forcing amplitude of each data record must be determined. 
We illustrated this procedure by considering two publicly available historical data records for pre-vaccination pertussis incidence. Based on these relatively short time series, we found that the output of the SIRSI model agrees with the main aspects of the phenomenology of the disease in a broad parameter range. It reproduces the multiannual interepidemic periods and the amplitude of fluctuations found in the data with a better agreement than SIR for the latter, while the SIRWS variant tends to produce stochastic fluctuations even larger than SIR.

Considerations about the recovery profile and its influence on the model's behavior concur in favoring models such as SIRSI, with a much flatter spectrum than SIRWS.
Here we have followed the common approach of taking recovery as a constant rate stochastic process, instead of more realistic unimodal recovery profiles that have been proposed and studied in the literature \cite{andrea,lloyd2001b}. For unforced models, it is known \cite{andrea} that the effect of this change is twofold, on one hand a small displacement of the main stochastic peak towards higher frequencies, and on the other an enhancement of the fluctuations. We have checked (results not shown) that this holds too for the SIRSI and SIRWS models in the parameter region explored in the previous section when we change from an exponentially distributed recovery profile to a unimodal gamma distributed recovery profile with the same average. Therefore, under more
realistic recovery profiles, the output of the SIRWS model
would be further away from the target power spectra, as would that
of the SIR model as well.

However, the short length of the time series used in our illustrative study prevents any definite
conclusion about the performance of the SIR model 
alternatives. Indeed, the ensemble of the
power spectra obtained from simulated time series of this length exhibits enough variability to
accommodate the two samples of real data. An extension of this analysis to a larger set of data could help to further elucidate the dynamics of disease acquired immunity, setting a solid ground for more complex models to deal with vaccine acquired immunity.

\section{Acknowledgements}
The authors thank Gabriela Gomes for helpful discussions.
Financial support from the Portuguese Foundation for Science and Technology (FCT) under Contract No. POCTI/ISFL/2/261 is gratefully acknowledged. The first author (G.R.) was also supported by FCT under Grants No. SFRH/BD/32164/2006 and No. SFRH/BPD/ 69137/2010, and by Calouste Gulbenkian Foundation under its Program "Stimulus for  Research". This research was also partially supported by the National Science Foundation under Grant No. NSF PHY05-51164. The data used in this paper are available online from the International Infectious Disease Data Archive (IIDDA) at http://iidda.mcmaster.ca.

\end{document}


\title{Modeling the long term dynamics of pre-vaccination pertussis\\Supplementary Material}
\author{Ganna Rozhnova\footnote{corresponding author\\a\_rozhnova@cii.fc.ul.pt}}
\affiliation{Centro de F{\'\i}sica da Mat\'eria Condensada and 
Departamento de F{\'\i}sica, Faculdade de Ci{\^e}ncias da Universidade de 
Lisboa, P-1649-003 Lisboa Codex, Portugal}
\affiliation{Kavli Institute for Theoretical Physics, Kohn Hall, University of California, Santa Barbara, CA 93106-4030}
\author{Ana Nunes}
\affiliation{Centro de F{\'\i}sica da Mat\'eria Condensada and 
Departamento de F{\'\i}sica, Faculdade de Ci{\^e}ncias da Universidade de 
Lisboa, P-1649-003 Lisboa Codex, Portugal}
\maketitle
\noindent{}\textbf{Supplementary Online Material Outline}
 
\noindent{}I. Deterministic counterparts of the SIR, SIRS, SIRIS, SIRWS and SIRSI models.\\
\noindent{}II. SIRIS model for $\beta_1\neq0.05$.\\
\noindent{}II. SIRWS model for $\kappa\neq20$.\\
\noindent{}IV. Analytical computation of the dominant frequency of the multiannual peak.\\
\noindent{}V. Estimation of $R_0$ based on the average age at primary infection.\\
\noindent{}VI. Analytical power spectrum of the stochastic SIR and SIRS models for $\beta_1=0$. 

\section{Deterministic counterparts of the SIR, SIRS, SIRIS, SIRWS and SIRSI models}
\subsection{SIR model}
In the infinite population limit, $N\to\infty$, the equations of the SIR model for the susceptible and infective densities are

\begin{eqnarray}
\label{dSIR1}
\dfrac{ds}{dt}&=&\mu\left(1-s\right)-\beta(t) si, \\
\label{dSIR2}
\dfrac{di}{dt}&=&\beta(t) si-(\delta+\mu)i,
\label{dSIR2}
\end{eqnarray}  
\\     
\noindent{}where

\begin{equation}
\label{infrate}
\beta(t)=\beta_0 (1+\beta_1 \cos 2\pi t).
\end{equation} 
\\
\noindent{}The density of recovered individuals is thus given by

\begin{equation}
\label{recs}
r=1-s-i.
\end{equation} 
\\
\noindent{}For $\beta_1=0$, seasonal forcing is absent and the system has a trivial ($i^*=0$, $s^*=1$) and a nontrivial fixed point. The latter can be expressed in terms of the basic reproductive ratio $R_0$ of a disease defined as
the average number of secondary cases generated by one infected individual in a fully susceptible
population during one infectious period:

\begin{equation}
\label{endsir}
i^*=\dfrac{\mu(R_0-1)}{\beta_0}, \ \ s^*=\dfrac{1}{R_0}, 
\end{equation}
\\
\noindent{}where 

\begin{equation}
\label{R0}
R_0=\dfrac{\beta_0}{\delta+\mu}.
\end{equation}
\\
\noindent{}Stability analysis shows that the trivial equilibrium is stable if $R_0<1$ while 
the endemic equilibrium (\ref{endsir}) is stable if $R_0>1$. This means that the critical value of $R_0$ given 
by Eq. (\ref{R0}) above which the disease is endemic is $1$.
\subsection{SIRS model}
In the infinite population limit, the equations of the SIRS model are

\begin{eqnarray}
\label{dSIRS}
\dfrac{ds}{dt}&=&\mu\left(1-s\right)-\beta(t) si+ \gamma r, \\
\label{dSIRS2}
\dfrac{di}{dt}&=&\beta(t) si-(\delta+\mu)i,
\end{eqnarray}  
\\ 
\noindent{}where $\beta(t)$ and $r$ are given by Eqs. (\ref{infrate}) and (\ref{recs}), respectively. 
For $\beta_1=0$ the non trivial equilibrium values $s^*$ and $i^*$ of the model satisfy 

\begin{equation}
\label{endsisrs}
i^*=\dfrac{(\gamma+\mu)(\delta+\mu)(R_0-1)}{\beta_0(\gamma+\delta+\mu)}, \ \ s^*=\dfrac{1}{R_0}, 
\end{equation}
\\
\noindent{}where $R_0$ is given by Eq. (\ref{R0}). 

Although the SIRS model is not usually formulated with explicit classes for the
naive susceptibles and those that become susceptible because of immunity waning, and for 
infectives with primary or subsequent infections, they can easily be introduced.
Considering the fraction $s_1$ of naive susceptibles  and the fraction $s_2$ of
 susceptibles that have been infected before, $s_2=s-s_1$,  Eq. (\ref{dSIRS}) for $\beta_1=0$  becomes
 
\begin{eqnarray}
\dfrac{ds_1}{dt}&=&\mu\left(1-s_1\right)-\beta_0 s_1 i, \nonumber \\
\dfrac{ds_2}{dt}&=& - \beta_0 s_2 i- \mu s_2 + \gamma (1-s_1-s_2-i),
\label{sisrsclass_s}
\end{eqnarray}
\\
\noindent{}yielding in equilibrium $s_1^*= \mu/(\beta_0 i^* + \mu)$. Similarly, 
from  Eq. (\ref{dSIRS2}), the  evolution equations for 
the fraction of primary infections $i_1$ and other infection episodes  $i_2$, $i_2 = i - i_1$,
are 

\begin{eqnarray}
\dfrac{di_1}{dt}&=&\beta_0 s_1 i -(\delta+\mu)i_1, \nonumber \\
\dfrac{di_2}{dt}&=&\beta_0 s_2 i -(\delta+\mu)i_2, 
\label{sisrsclass_i}
\end{eqnarray}
\\
\noindent{}yielding in equilibrium $i_1^*/i^*= \beta_0 s_1^*/(\delta + \mu)= \mu /\left[\gamma (1-s^*-i^*) + \mu\right]$ for the fraction of primary infections.
\subsection{SIRWS model}
In the infinite population limit, the equations of the SIRWS model for the densities of individuals in the susceptible, infectious and waning classes are

\begin{eqnarray}
\label{dSIRWS1}
\dfrac{ds}{dt}&=&\mu\left(1-s\right)-\beta(t) si+2\gamma w, \\
\label{dSIRWS2}
\dfrac{di}{dt}&=&\beta(t) si-(\delta+\mu)i,\\
\label{dSIRWS3}
\dfrac{dw}{dt}&=&2\gamma(r-w)-\kappa\beta_0 w i-\mu w,
\label{dSIRWS3}
\end{eqnarray}
\\
\noindent{}where $r$ is the density of individuals in the recovered class

\begin{equation}
\label{rpnas}
r=1-s-i-w
\end{equation} 
\\
\noindent{}and $\beta(t)$ is given by Eq. (\ref{infrate}).  
\subsection{SIRIS model}
In the infinite population limit, the equations of the 
SIRIS model for the susceptible and infective densities are

\begin{eqnarray}
\label{gabriela1}
\dfrac{ds}{dt}&=&\mu\left(1-s\right)-s\left[\beta(t) i_1+\beta_0\eta i_2\right]+\gamma r, \\
\dfrac{di_1}{dt}&=&s\left[\beta(t) i_1+\beta_0\eta i_2\right]-(\delta+\mu)i_1,\\
\label{gabriela2}
\dfrac{di_2}{dt}&=&\sigma\beta_0 r\left(i_1+\eta i_2\right)-(\delta+\mu)i_2,
\label{gabriela3}
\end{eqnarray}
\\
\noindent{}where

\begin{equation}
\label{recsgabriela}
r=1-s-i_1-i_2
\end{equation} 
\\
\noindent{}and $\beta(t)$ is given by Eq. (\ref{infrate}). If $\beta_1=0$ the model has a trivial equilibrium and an endemic equilibrium where the densities of infectives and susceptibles lie between 0 and 1. The analytical computation of the endemic equilibrium values of $i_1^*$, $i_2^*$ and $s^*$ is straightforward but the expressions are lengthy and that is why we do not write them explicitly.
\subsection{SIRSI model}
In the infinite population limit, the equations of the 
SIRSI model for the susceptible and infective densities are

\begin{eqnarray}
\label{dS1I1RS2I2R}
\dfrac{ds_1}{dt}&=&\mu\left(1-s_1\right)-s_1\left[\beta(t) i_1+\beta_0\eta i_2\right], \\
\dfrac{di_1}{dt}&=&s_1\left[\beta(t) i_1+\beta_0\eta i_2\right]-(\delta+\mu)i_1,\\
\dfrac{ds_2}{dt}&=&\gamma r-\mu s_2-\sigma\beta_0s_2\left(i_1+\eta i_2\right),\\
\label{dS1I1RS2I2R2}
\dfrac{di_2}{dt}&=&\sigma\beta_0s_2\left(i_1+\eta i_2\right)-(\delta+\mu)i_2,
\end{eqnarray} 
\\
\noindent{}where

\begin{equation}
\label{recsrohani}
r=1-s_1-i_1-s_2-i_2
\end{equation} 
\\
\noindent{}and $\beta(t)$ is given by Eq. (\ref{infrate}). It can be shown that for $\beta_1=0$ the model has a trivial equilibrium and an endemic equilibrium where the densities of infectives and susceptibles lie between 0 and 1. As for the previous model, we do not give its explicit form. 

A comparison of the equilibrium infective densities corresponding to the nontrivial fixed point as a function of $\beta_0$ for the five models is shown in Fig. 2 of the main text.

\section{SIRIS model for $\beta_1\neq0.05$}
\begin{figure}
\includegraphics[trim=0cm 0cm 0cm 0cm, clip=true, width=0.3333\textwidth]{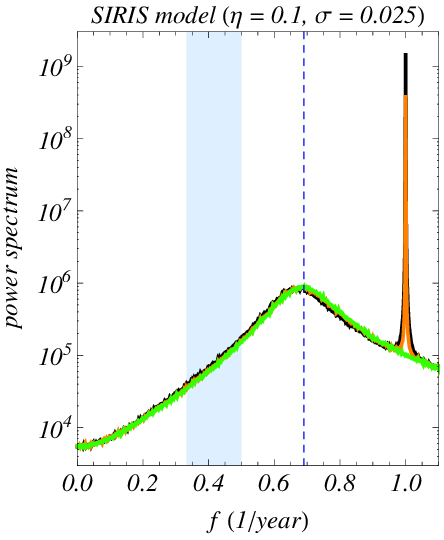}
\caption{(Color online) The dependence of the power spectrum of the SIRIS model on the amplitude of seasonal forcing $\beta_1$. Parameters: $\beta_1=0 \text{ [green (light gray)], } 0.05 \text{ [orange (dark gray)], and } 0.1 \text{ (black)}$. The rest of the parameter values as for the black line in the left panel of Fig. 7 of the main text.}
\label{fig2a}
\end{figure}

In the main text we presented the results for the SIRIS model for fixed $\beta_1=0.05$. Figure \ref{fig2a} shows the power spectra of the number of infective individuals from simulation for $\beta_1=0,0.05,0.1$. The stochastic peaks agree perfectly for all $\beta_1\in[0,0.1]$ and they are always far away from the shaded region. There is a remarkable resemblance of these spectra with those of the SIRS model for the same value of immunity waning rate (compare them with the black lines in Fig. 6 of the main text).

\begin{figure}
\includegraphics[trim=0cm 0cm 0cm 0cm, clip=true, width=\textwidth]{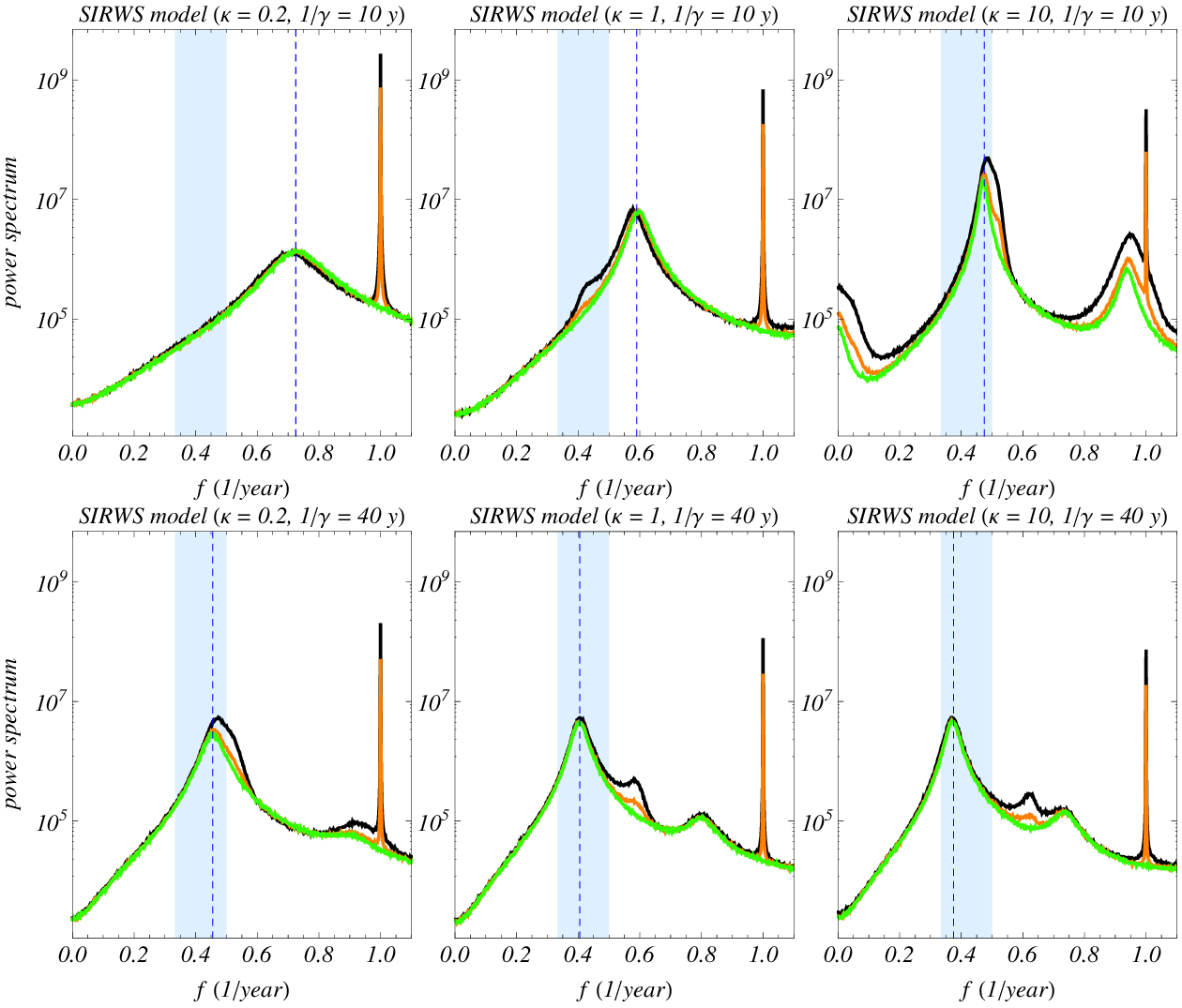}
\caption{(Color online) The dependence of the power spectrum of the SIRWS model on the boosting coefficient $\kappa$. Parameters: $\beta_1=0 \text{ [green (light gray)], } 0.05 \text{ [orange (dark gray)], and } 0.1 \text{ (black)}$.}
\label{fign}
\end{figure}

\section{SIRWS model for $\kappa\neq20$}

In the main text we presented the results for the SIRWS model for fixed boosting coefficient $\kappa=20$. Figure \ref{fign} shows the power spectra of the number of infective individuals from simulation for $\kappa=0.2,1,10$ and $1/\gamma=10$ years (top panels), 40 years (bottom panels). In each panel there are three spectra corresponding to different values of the seasonal forcing amplitude. For given $\gamma$ and $\kappa$ the dominant stochastic peaks agree well for all $\beta_1\in[0,0.1]$. We also observe that they are situated within the shaded region unless in the limit of the SIRS model ($\kappa$ is small and $\gamma\gg \mu$).

\section{Analytical computation of the dominant frequency of the multiannual peak}
\label{analytpeak}
In this section we will describe how the dominant frequencies of the multiannual stochastic peaks in power spectra from simulations can be computed analytically. First, we note that in the three most interesting models which are the SIRWS, the SIR and the SIRSI model these frequencies are independent of $\beta_1 \in [0,0.1]$ (see Fig. 5, Fig. 8 and Fig. 9 of the main text). Thus, to predict the position of the stochastic peak in this parameter range it is sufficient to consider the unforced models with $\beta_1=0$ whose analysis is easier. The method we are going to use is usually referred to as the system size expansion first developed by van Kampen \cite{vankampen}. It has been successfully applied to several epidemiological models based on the S(E)IR or SIRS dynamics. We refer the reader to \cite{mckane1,PhysRevE.79.041922,mckanewcough} for a more extensive discussion of the method in the context of unforced epidemiological models. The analysis of seasonally forced models can be found in \cite{seasonal,forcedmckane}. Here we will outline the most important steps of the calculations for the unforced SIRSI model because it has the largest number of classes.  

The analytical treatment of the stochastic SIRSI model simulated using Gillespie's algorithm starts with the construction of the master equation \cite{vankampen}. We describe the state of the system $l\equiv(l_1,l_2,l_3,l_4)$ by four numbers which are, sequentially, the number of susceptible $S_1$, infective $I_1$, susceptible $S_2$, and infective $I_2$ individuals. The total population size $N$ is fixed, that is why the number of recovered individuals $R$ at any time is given by $N-l_1-l_2-l_3-l_4$. The transition rates $\mathcal{T}^{l'}_l$ from the initial state $l$ to the final state $l'$ associated to the processes postulated in the diagram (e) of Fig. 1 of the main text can be written as follows:
\begin{enumerate}
	\item Immunity waning: $R\stackrel{\gamma}{\rightarrow} S_2$. 
\begin{equation}
\label{RtoS2}
\mathcal{T}^{l_1,l_2,l_3+1,l_4}_{l_1,l_2,l_3,l_4}=\gamma(N-l_1-l_2-l_3-l_4). 
\end{equation}
	\item Infection: $S_1\stackrel{\bar{\lambda}_c}{\longrightarrow} I_1$ and $S_2\stackrel{\sigma\bar{\lambda}_c}{\longrightarrow} I_2$, where $\bar{\lambda}^c=\beta_0(I_1+\eta I_2)/N$. 
\begin{equation}
\label{S1toI1}
\mathcal{T}^{l_1-1,l_2+1,l_3,l_4}_{l_1,l_2,l_3,l_4}=\beta_0l_1(l_2+\eta l_4)/N,
\end{equation}
\begin{equation}
\label{S2toI2}
\mathcal{T}^{l_1,l_2,l_3-1,l_4+1}_{l_1,l_2,l_3,l_4}=\sigma\beta_0l_3(l_2+\eta l_4)/N. 
\end{equation}
  \item Recovery: $I_1,I_2\stackrel{\delta}{\rightarrow} R$.  
\begin{equation}
\label{I1toR}
\mathcal{T}^{l_1,l_2-1,l_3,l_4}_{l_1,l_2,l_3,l_4}=\delta l_2, 
\end{equation}
\begin{equation}
\label{I2toR}
\mathcal{T}^{l_1,l_2,l_3,l_4-1}_{l_1,l_2,l_3,l_4}=\delta l_4. 
\end{equation}
  \item Birth and death: $R\stackrel{\mu}{\rightarrow} S_1$ and $S_1,I_1,S_2,I_2\stackrel{\mu}{\rightarrow} R$.
\begin{equation}
\label{RtoS1}
\mathcal{T}^{l_1+1,l_2,l_3,l_4}_{l_1,l_2,l_3,l_4}=\mu N,
\end{equation}
\begin{equation}
\label{S1toR}
\mathcal{T}^{l_1-1,l_2,l_3,l_4}_{l_1,l_2,l_3,l_4}=\mu l_1, 
\end{equation}
\begin{equation}
\label{S2toR}
\mathcal{T}^{l_1,l_2-1,l_3,l_4}_{l_1,l_2,l_3,l_4}=\mu l_2, 
\end{equation}
\begin{equation}
\label{I1toR}
\mathcal{T}^{l_1,l_2,l_3-1,l_4}_{l_1,l_2,l_3,l_4}=\mu l_3,
\end{equation}
\begin{equation}
\label{I2toR}
\mathcal{T}^{l_1,l_2,l_3,l_4-1}_{l_1,l_2,l_3,l_4}=\mu l_4. 
\end{equation}
\end{enumerate}

The master equation for the probability $\mathcal{P}(l,t)$ of having the system in state $l$ at time $t$ in the SIRSI model can be obtained by substituting Eqs. (\ref{RtoS2})-(\ref{I2toR}) into the general form for the master equation \cite{vankampen}:

\begin{equation}
\label{generalmasterequation}
\frac{d\mathcal{P}(l,t)}{dt}=\sum\limits_{l'\neq l} \left[\mathcal{T}^l_{l'}\mathcal{P}(l',t)-\mathcal{T}^{l'}_l\mathcal{P}(l,t) \right].
\end{equation}
\\
\noindent{}Given the initial and boundary conditions, the solution of Eq. (\ref{generalmasterequation}) is equivalent to the full stochastic simulation of the model. The exact analytical solution of the master equation for the SIRSI model is not feasible but we can study its limit when the size of the system, $N$, becomes large \cite{vankampen}. The limit is formalized by representing the discrete variables $l_k$ as a sum of two terms: 

\begin{equation}
l_k(t)=Nd_k(t)+\sqrt{N}x_k(t), \ \ \ k=1,\ldots,4.
\end{equation}
\\
\noindent{}The first term is the deterministic macroscopic term of the order of $N$ and the second term are the fluctuations of the order of $\sqrt{N}$ around the macroscopic state. To preserve the notation introduced in the main text we denote the functions $d_k(t)$

\begin{equation}
\label{vk}
d_k(t)=\lim_{N\to\infty}\frac{l_k(t)}{N}
\end{equation}
\\
\noindent{}as the densities of susceptible individuals $s_1(t)$ for $k=1$ and $s_2(t)$ for $k=3$, and of infective individuals $i_1(t)$ for $k=2$ and $i_2(t)$ for $k=4$.

\begin{figure}
\includegraphics[trim=0cm 0cm 0cm 0cm, clip=true, width=0.3333\textwidth]{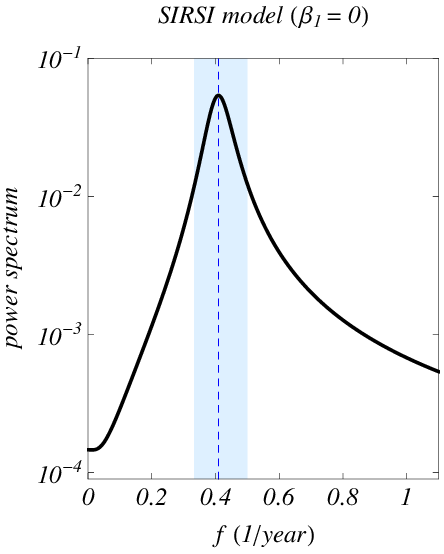}
\caption{(Color online) Analytical power spectrum of the unforced SIRSI model for parameter values used in the right panel of Fig. 9 of the main text. The shaded region marks the frequencies between 1/3 $\text{year}^{-1}$ and 1/2 $\text{year}^{-1}$ and the dashed blue vertical line corresponds to the frequency of the maximum of the spectrum.}
\label{figs7}
\end{figure}

Substituting Eq. (\ref{vk}) into Eq. (\ref{generalmasterequation}) results in an equation for the probability distribution $\Pi(x,t)$ of the fluctuations $x\equiv(x_1,x_2,x_3,x_4)$ around the deterministic trajectory in which terms of different orders in $1/\sqrt{N}$ can be identified. After collecting and equating terms of the leading order one obtains the deterministic SIRSI equations with $\beta_1=0$ [see Eqs. (19)-(22)]. The same procedure for terms of the next to leading order gives rise to a linear Fokker-Plank equation in the variable $x$ \cite{vankampen}:

\begin{equation}
\label{fokkerplanckseas}
\dfrac{\partial\Pi(x,t)}{\partial t}=-\sum\limits_{k,j}{A_{kj}(t)\dfrac{\partial (x_j\Pi(x,t))}{\partial x_k}}+\frac{1}{2}\sum\limits_{k,j}{B_{kj}(t)\dfrac{{\partial}^2\Pi(x,t)}{\partial x_k \partial x_j}}, \ \ \ k,j=1,\ldots,4.
\end{equation}
\\         
\noindent{}In Eq. (\ref{fokkerplanckseas}), ${\bf A}(t)$ is the Jacobian of the deterministic SIRSI model with $\beta_1=0$
 
\begin{equation}
\label{Aseir}
{\bf A}(t)= \left( \begin{array}{cccc} -g-\mu & -\beta_0 s_1 & 0 & -\beta_0 \eta s_1 \\
g & \beta_0 s_1-(\delta+\mu) & 0 & \beta_0 \eta s_1 \\
-\gamma & -\sigma\beta_0 s_2 -\gamma &- \sigma g-(\gamma+\mu) & -\sigma\beta_0\eta s_2-\gamma  \\
0 & \sigma\beta_0 s_2 &  \sigma g & \sigma\beta_0\eta s_2-(\delta+\mu) \end{array} \right) 
\end{equation}
\\
\noindent{}and ${\bf B}(t)$ is the symmetric cross correlation matrix computed directly from the expansion

\begin{equation}
\label{Bseir}
{\bf B}(t)= \left( \begin{array}{cccc} gs_1+\mu(1+s_1) & -g s_1 & 0  & 0 \\
-g s_1 & gs_1+ (\delta+\mu)i_1 & 0 & 0\\
0 & 0 & \gamma r +(\sigma g+\mu) s_2& -\sigma gs_2\\
0 &  0 & -\sigma gs_2 & \sigma g s_2+(\delta+\mu)i_2 \end{array} \right) ,  
\end{equation}
\\
\noindent where 
  
\begin{equation}
g=\beta_0(i_1+\eta i_2)
\end{equation}
\\
\noindent{}and
 
\begin{equation}
r=1-s_1-i_1-s_2-i_2.
\end{equation}
\\
To compute the power spectrum of the fluctuations it is useful to use the Langevin equation \cite{vankampen} associated with Eq. (\ref{fokkerplanckseas}) which reads as

\begin{equation}
\label{langevinseir}
\dfrac{dx_k(t)}{dt}=\sum\limits_j{A_{kj} (t) x_j(t)}+L_k(t), \ \ k,j=1,\ldots,4,
\end{equation}
\\
\noindent{}where $L_k(t)$ are Gaussian noise terms with zero mean 

\begin{equation}
          \left\langle L_k(t)\right\rangle =0 
\end{equation}
\\
\noindent{}and with the correlator

\begin{equation}
\left\langle L_k(t)L_j(t')\right\rangle =B_{kj}(t)\delta(t-t') .
\end{equation}
\\
Finally, since we are interested in the fluctuations in the stationary state we evaluate the matrices ${\bf A}(t)$ and ${\bf B}(t)$ at the endemic steady state $(s_1^*,i_1^*,s_2^*,i_2^*)$ of the deterministic SIRSI equations (with $\beta_1=0$). The power spectrum of the fluctuations for infectives $I_1$ around the endemic equilibrium of the unforced SIRSI model is defined as $P_{I_1}(\omega)\equiv\left\langle|\tilde x_2(\omega)|^2\right\rangle$. Figure \ref{figs7} shows $P_{I_1}$ as a function of frequency measured in the units of $f=\omega/2\pi$ for the parameter values used in Fig. 9 of the main text. As before, the blue shaded region corresponds to multi-annual periods in the range 2 to 3 years where the multiannual peaks of the spectra computed from London and Ontario data are located. The frequency corresponding to the center of the shaded region, $f_c$, is thus $5/12$ $\text{year}^{-1}$. The frequency of the maximum of the spectrum, $f_{peak}$, is $0.409401$ $\text{year}^{-1}$. We used this value to draw the dashed blue vertical line in the right panel of Fig. 9 of the main text. 

\begin{figure}
\includegraphics[trim=0cm 0cm 0cm 0cm, clip=true, width=0.49\textwidth]{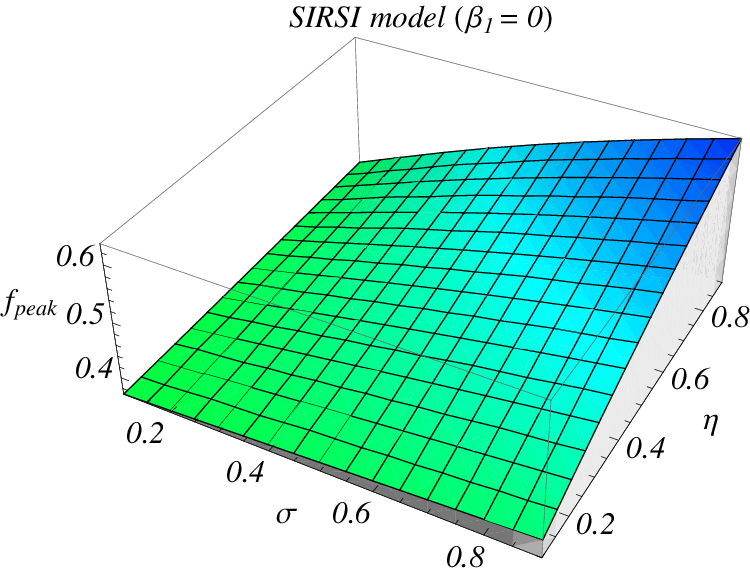}
\caption{(Color online) The frequency of the maximum of the analytical power spectrum of the unforced SIRSI model as a function of $\eta$ and $\sigma$. The plot is made for $1/\gamma=20$ years. These values were used to draw dashed vertical lines in Fig. \ref{figs7} of this section and in Fig. 9 of the main text.}
\label{figs8}
\end{figure}

As described in the main text we also studied the properties of the spectra for the unforced SIRSI model for other parameter values. For fixed $\gamma$, we considered in the interval $(0,1)$ nine equally spaced values and constructed a grid of 81 points in the $(\eta,\sigma)$ space. Figure \ref{figs8} shows the values of $f_{peak}$ on this grid for $1/\gamma=20$ years. We further calculated the number of spectra $N_{target}$ for which
 
\begin{equation}
\frac{1}{3}\leq f_{peak}\leq\frac{1}{2}             
\end{equation}
\\
\noindent{}and the number of spectra $N_0$ for which 
\begin{equation}
\frac{f_c}{1.05}\leq f_{peak}\leq \frac{f_c}{0.95}. 
\end{equation}
\\
\noindent{}The former gives the total number of spectra in the target region, and the latter gives the number of spectra with periods within 5\% of the period corresponding to the center of the target region. The results are summarized in Table \ref{tableS} for three different values of $\gamma$. 

\begin{figure}
\includegraphics[trim=0cm 0cm 0cm 0cm, clip=true, width=0.49\textwidth]{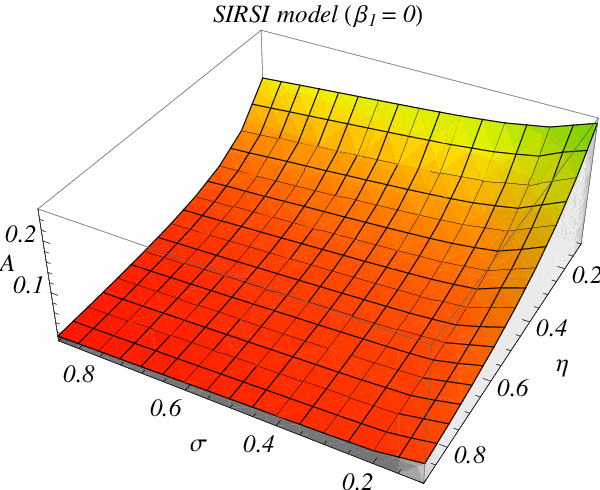}
\caption{(Color online) The amplitude $A$ of the peak of the analytical power spectrum 
of the unforced SIRSI model as a function of $\eta$ and $\sigma$. The plot is made for $1/\gamma=20$ years.}
\label{figs9}
\end{figure}

\begin{table}
\centering
{\begin{tabular}{cccc}
\hline\hline \ \ \ $\gamma$ ($\text{year}^{-1}$) \ \ \ & $1/40$  &  $1/20$  &  $1/10$  \\  
\hline $N_{target}$ & 78 & 62 & 43 \\
\hline $N_0$ & 28 & 20 & 16 \\ 
\hline \ \ \ $A_{min}$ \ \ \ & \ \ \ 0.0602494 \ \ \ & \ \ \ 0.0280422 \ \ \ & \ \ \ 0.0159649 \ \ \ \\
\hline $A_{max}$ & 0.165027 & 0.166171 & 0.123987 \\
\hline\hline
\end{tabular}
\caption{Values for the quantities used in the comparison and description of the spectra for the unforced SIRSI model in the $(\eta,\sigma)$ space and different values of $\gamma$.}
\label{tableS}}
\end{table}

Another quantity of interest in the description of power spectra is the amplitude of the peak $A$. Figure \ref{figs9} shows the amplitude for the same 81 points with $\eta,\sigma\in (0,1)$ and $1/\gamma=20$ years. The value of $A$ is particularly interesting for the $N_0$ spectra with their peaks in center of the target region. In Table \ref{tableS} we give the maximum and the minimum amplitude among these $N_{0}$ spectra for different values of $\gamma$. The amplitude of the peak plotted in Fig. \ref{figs7} of this section is 0.0538342, in the middle of the range of $A$ when $1/\gamma=20$ years.   

This study of the unforced SIRSI model is applicable to the case $\beta_1\neq0$, because we observe that the amplitude of the stochastic peak does not change for the seasonally forced SIRSI model.

The conclusion of this section is that the SIRSI model's spectrum is robust with respect to variation of all free parameters. On one hand, there is a large region of parameter space where the dominant frequency of the stochastic peak is centered in the region of interest. On the other hand, these spectra have some variability in the peak's amplitude for different values of $\eta$, $\sigma$ and $\gamma$.

\section{Estimation of $R_0$ based on the average age at primary infection}
The results for all models presented in the main text are obtained for fixed basic reproductive ratio $R_0=17$. Here we discuss the effect of relaxation of this assumption. 

\subsection{SIRSI model}
We first consider the SIRSI model as an example. Under the assumption of permanent immunity, $R_0$ can be estimated from the average age at primary infection, $\mathcal{A}$, and the average lifespan, $1/\mu$, as $R_0=1/(\mu \mathcal{A})$ \cite{amay}. Thus the parameter values used in the main text (see Table I of the main text) correspond to the average age at primary infection $\mathcal{A}=50/17\approx2.94$ years. On the other hand, using Eqs. (\ref{dS1I1RS2I2R}), $\mathcal{A}$ can be found from the relation $\mathcal{A}=1/\left[\mu+ \bar{\lambda}_t\right]$ (for notation see Fig. 1 in the main text). Here $\bar{\lambda}_t=\left(\beta(t)i_1+\beta_0 \eta i_2\right)$ is the rate of infection of naive individuals and $\mu$ is the death rate of individuals. Evaluating this expression in the absence of seasonality, $\beta_1=0$, at the endemic fixed point gives $\mathcal{A}=1/\left[\mu+ \beta_0\left(i_1^*+\eta i_2^*\right)\right]$. From this equation we can obtain the average contact rate, $\beta_0$, corresponding to the average age at infection of naive individuals estimated as $50/17$ years. To achieve this, we use the analytical expressions for $i_1^*$ and $i_2^*$ where we substitute the numerical values for all parameters but $\beta_0$ and solve $50/17=1/[\mu+\beta_0\left(i_1^*+\eta i_2^*\right)]$ for $\beta_0$.

Using for a fixed value of $\gamma$ the same grid of 81 points in the $(\eta,\sigma)$ space we finally recompute the maximum and minimum values for $R_0=\beta_0/(\gamma+\mu)$ on this grid and the number of target spectra with the peaks situated in the shaded region, $N_{target}$. The results are summarized in Table II.

\begin{table}
\centering
{\begin{tabular}{cccc}
\hline\hline \ \ \  $\gamma$ ($\text{year}^{-1}$) \ \ \  & \ \ \ \ \ \ $1/40$ \ \ \ \ \ \ & \ \ \ \ \ \ $1/20$ \ \ \ \ \ \ & \ \ \ \ \ \ $1/10$ \ \ \ \ \ \ \\  
\hline $N_{target}$ & 53 & 41 & 30 \\
\hline ${R_0}_{min}$ & 8.63 & 6.06 & 4.09 \\ 
\hline ${R_0}_{max}$ & 16.16 & 15.77 & 15.38 \\ 
\hline\hline
\end{tabular}
\caption{Number of the spectra in the target region, $N_{target}$, and the minimum and the maximum values of $R_0$ computed on the grid of 81 points in the $(\eta,\sigma)$ space.}
\label{tableSS}}
\end{table}

From comparison of the results shown in Table I and Table II we observe that the number of spectra in the target region reduces slightly if the assumption of fixed $R_0$ is relaxed. However, we see that even for the largest value of $\gamma$ we explored at least 37\% of the spectra correspond to the interepidemic periods compatible with the London and Ontario data. We conclude that the results presented for the model in the main text are very robust with respect to the parametrization of the model.

\subsection{SIRS model}

A similar analysis can be carried out for the SIRS model, solving for $\beta_0$ the equation $50/17=1/\left[\mu+\beta_0 i^*\right]$ for each parameter choice. For the range of average immunity period explored in the main text, $1/\gamma\in[10,40]$ years, this approach yields $R_0=\beta_0/(\gamma+\mu)$ in the range $[3.68,8.12]$ and all the stochastic peaks are shifted to the left with respect to those shown in Fig. 6 of the main text. More specifically, we observe that for the whole range of $R_0$ the peak's frequencies in the SIRS model are equal to that in the SIR model, $f_{peak}=0.37$ $\text{year}^{-1}$. We conclude that under the assumption that the totally naive population is well mixed, the SIRS model fails to produce spectra with the stochastic peak's frequencies significantly different from that in the SIR model.

\subsection{SIRWS and SIRIS models}

We finally give the expressions for the SIRWS and the SIRIS models from where $R_0$ keeping the average age at first infection fixed at its value for the SIR model can be computed for each parameter choice. For the former, $\beta_0$ is found from $50/17=1/\left[\mu+\beta_0 i^*\right]$, where $i^*$ is the density of infectives in equilibrium of Eqs. (\ref{dSIRWS1})-(\ref{dSIRWS3}). For the latter, we solve $50/17=1/\left[\mu+\beta_0 \left(i_1^*+\eta i_2^*\right)\right]$ for $\beta_0$, where $i_1^*$ and $i_2^*$ are the densities of individuals with severe and mild infections in equilibrium of Eqs. (\ref{gabriela1})-(\ref{gabriela3}).

\section{Analytical power spectrum of the stochastic SIR and SIRS models for $\beta_1=0$}

The analytical power spectrum of the fluctuations for infectives $I$ around the endemic equilibrium of the unforced ($\beta_1=0$) SIR, Eqs. (\ref{dSIR1})-(\ref{dSIR2}), and SIRS, Eqs. (\ref{dSIRS})-(\ref{dSIRS2}), models can be computed following the approach described in Section \ref{analytpeak} of the Supplementary Material. For each model, this analytical expression can be used to predict the dominant frequency of the stochastic peak in the corresponding stochastic seasonally-forced model if the position of the peak does not depend or depends only very slightly on $\beta_1$. This is typically the case when $\beta_1$ is not very large so that the stochastic model exhibits fluctuations around the period one stable attractor.

\subsection{SIR model}    

The power spectrum for the fluctuations of infectives in the unforced SIR model is given by

\begin{equation}
P_I(\omega)=\dfrac{2\mu (\beta_0-\delta-\mu)}{Z\beta_0}\left(\omega^2+\dfrac{(\mu\beta_0)^2}{(\delta+\mu)^2}\right),
\end{equation}
\\
\noindent{}where $Z=(D-\omega^2)^2+T^2\omega^2$, $D=\mu(\beta_0-\delta-\mu)$ and $T=-\beta_0\mu/(\delta+\mu)$. Here $D$ and $T$ are the determinant and the trace of the linear approximation of Eqs. (\ref{dSIR1})-(\ref{dSIR2}) at the endemic equilibrium, Eq. (\ref{endsir}). We refer the reader to Ref. \cite{MSimoes05062008} for details on the computation of the SIR power spectrum.

\subsection{SIRS model}    

The power spectrum for the fluctuations of infectives in the unforced SIRS model is given by

\begin{equation}
P_I(\omega)=\dfrac{2(\delta+\mu)(\gamma+\mu)(\beta_0-\delta-\mu)}{Z\beta_0 (\gamma+\delta+\mu)}\left(\omega^2+\dfrac{K(\gamma+\mu)}{(\gamma+\delta+\mu)^2}\right),
\end{equation}
\\
\noindent{}where $Z=(D-\omega^2)^2+T^2\omega^2$, $D=(\gamma+\mu)(\beta_0-\delta-\mu)$, $T=-(\gamma+\mu)(\beta_0+\gamma)/(\delta+\gamma+\mu)$ and $K=(\beta_0+\gamma)^2(\gamma+\mu)-\gamma(\beta_0-\delta-\mu)(\gamma+\delta+\mu)$. Here $D$ and $T$ are the determinant and the trace of the linear approximation of Eqs. (\ref{dSIRS})-(\ref{dSIRS2}) at the endemic equilibrium, Eq. (\ref{endsisrs}). We refer the reader to Refs. \cite{PhysRevE.79.041922} and \cite{PhysRevE.80.051915} for details on the computation of the SIRS power spectrum.